\documentclass[conference]{IEEEtran}
\IEEEoverridecommandlockouts
\usepackage{cite}
\usepackage{amsmath,amssymb,amsfonts}
\usepackage{algorithmic}
\usepackage{graphicx}
\usepackage{subcaption}
\usepackage{tabularx}
\usepackage{booktabs}
\usepackage{url}
\usepackage{textcomp}
\usepackage{xcolor}
\usepackage{multirow}
\usepackage{diagbox}

\def\BibTeX{{\rm B\kern-.05em{\sc i\kern-.025em b}\kern-.08em
    T\kern-.1667em\lower.7ex\hbox{E}\kern-.125emX}}
\begin{document}

\title{SPH-Net: A Co-Attention Hybrid Model for Accurate Stock Price Prediction\\
\thanks{$^*$ Equal Contribution; $^{\dag}$ Corresponding Author}
}
\author{
\IEEEauthorblockN{Yiyang Wu$^{1,*}$, Hanyu Ma$^{2,*}$, Muxin Ge$^{3}$, Xiaoli Ma$^{4}$, Yadi Liu$^{5}$, Ye Aung Moe$^{6}$, Zeyu Han$^{3,\dag}$, Weizheng Xie$^{3,\dag}$}
\IEEEauthorblockA{\textit{$^{1}$WuXpress Warehousing LLC, $^{2}$Walnut Grove High School, $^{3}$Southern Methodist University} \\
\textit{$^{4}$Washington State University, $^{5}$Nanyang Technological University, $^{6}$University of Nebraska-Lincoln}\\
mark2789129293@gmail.com$^{*}$, weizhengx@smu.edu$^{\dag}$}
}

\maketitle

\begin{abstract}
Prediction of stock price movements presents a formidable challenge in financial analytics due to the inherent volatility, non-stationarity, and nonlinear characteristics of market data. This paper introduces SPH-Net (Stock Price Prediction Hybrid Neural Network), an innovative deep learning framework designed to enhance the accuracy of time series forecasting in financial markets. The proposed architecture employs a novel co-attention mechanism that initially processes temporal patterns through a Vision Transformer, followed by refined feature extraction via an attention mechanism, thereby capturing both global and local dependencies in market data. To rigorously evaluate the model's performance, we conduct comprehensive experiments on eight diverse stock datasets: AMD, Ebay, Facebook, FirstService Corp, Tesla, Google, Mondi ADR, and Matador Resources. Each dataset is standardized using six fundamental market indicators: Open, High, Low, Close, Adjusted Close, and Volume, representing a complete set of features for comprehensive market analysis. Experimental results demonstrate that SPH-Net consistently outperforms existing stock prediction models across all evaluation metrics. The model's superior performance stems from its ability to effectively capture complex temporal patterns while maintaining robustness against market noise. By significantly improving prediction accuracy in financial time series analysis, SPH-Net provides valuable decision-support capabilities for investors and financial analysts, potentially enabling more informed investment strategies and risk assessment in volatile market conditions.

\end{abstract}


\section{Introduction}

Stock price prediction represents a fundamental yet challenging research domain in computational finance, owing to the inherent volatility, non-linearity, and stochastic nature of financial markets \cite{bg-1}. While traditional statistical approaches, including ARIMA models \cite{spp-arima} and econometric frameworks \cite{spp-kappa}, have established foundational methodologies for time series analysis, their capacity to capture complex market dynamics and long-range dependencies remains limited \cite{lstm,att}.

The advent of artificial intelligence, particularly deep learning, has catalyzed significant advancements in financial forecasting. Machine learning algorithms such as Support Vector Machines \cite{svm}, Random Forests \cite{rf,rf1}, and Gradient Boosting methods \cite{gg} have demonstrated improved performance by modeling non-linear relationships in market data. Subsequent developments in recurrent neural architectures, including RNNs \cite{rnn} and their advanced variants (LSTMs \cite{lstm} and GRUs \cite{gru}), have further enhanced temporal pattern recognition capabilities.

The recent emergence of Transformer-based models \cite{att} has introduced new possibilities through their attention mechanisms, which effectively capture global contextual relationships across extended time sequences. However, current approaches still exhibit limitations in processing heterogeneous financial data and adapting to real-world market volatility \cite{att,vit}.

To overcome these challenges, we present SPH-Net (\textbf{S}tock \textbf{P}rice \textbf{H}ybrid Neural \textbf{Net}work), an innovative hybrid architecture that synergistically combines Vision Transformer (ViT) \cite{vit} with Transformer encoder-decoder mechanisms \cite{att}. Our framework employs co-attention mechanisms to simultaneously model temporal dependencies and cross-feature interactions in financial time series data.

The principal contributions of this work are threefold:
\begin{itemize}
    \item A novel hybrid deep learning architecture that integrates ViT and Transformer components for comprehensive stock price prediction
    \item An innovative data preprocessing methodology that transforms temporal financial data into image-like patch representations, facilitating enhanced pattern extraction through ViT
    \item Comprehensive empirical validation across eight diverse stock datasets, demonstrating SPH-Net's superior performance over state-of-the-art models in both regression and classification scenarios
\end{itemize}

\section{Related Work}

\subsection{Contemporary AI Models for Stock Price Forecasting}

Early approaches to stock price prediction predominantly employed classical statistical methods, including Autoregressive Integrated Moving Average (ARIMA) models \cite{spp-arima} and linear regression techniques \cite{lir,lor,spp-kappa}. While these methods established foundational frameworks for time series analysis, their inherent linearity constrained their capacity to model the complex, non-linear dynamics characteristic of financial markets.

\begin{figure*}[h]
  \centering
  \includegraphics[width=1.8\columnwidth]{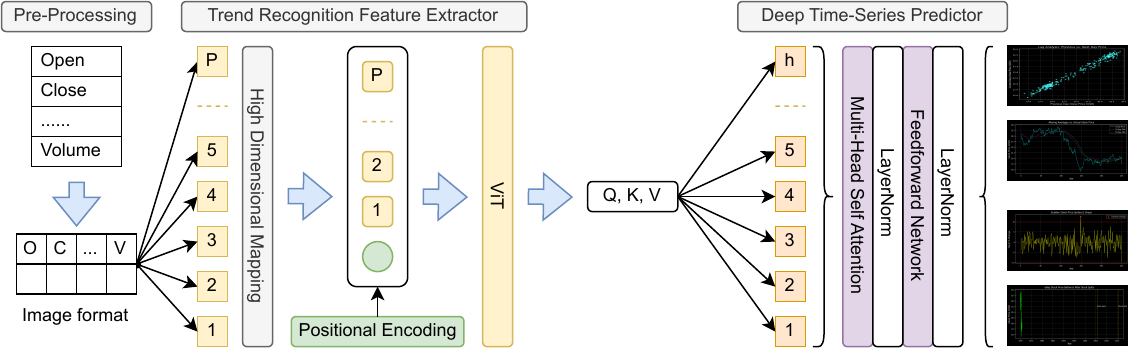}
  \caption{Architectural overview of the proposed SPH-Net framework.} 
  \label{stru}
\end{figure*}

The emergence of machine learning paradigms introduced more sophisticated algorithms, notably Support Vector Machines (SVMs) \cite{svm}, Decision Trees \cite{dt}, and Random Forests \cite{rf}. These techniques demonstrated improved predictive performance through enhanced pattern recognition capabilities. Nevertheless, their effectiveness remained limited when modeling sequential dependencies in temporal data.

Recent advances in deep learning have revolutionized the field, with Recurrent Neural Networks (RNNs) \cite{rnn}, Long Short-Term Memory networks (LSTMs) \cite{lstm,bibm}, Gated Recurrent Units (GRUs) \cite{gru}, and hybrid CNN-LSTM architectures \cite{cnn-lstm} demonstrating superior temporal modeling performance. Specifically, LSTM and GRU variants excel at capturing long-range dependencies in financial time series, while CNN-LSTM hybrids benefit from synergistic spatial-temporal feature extraction.

The introduction of Transformer-based architectures \cite{att,bert,vit,cmin,in-1} has marked a significant paradigm shift, with self-attention mechanisms enabling more effective modeling of long-range dependencies. Notably, BERT \cite{bert} has been adapted for time-series analysis, though its conventional architecture lacks the encoder-decoder structure typically required for forecasting applications.

\subsection{Limitations of Current Approaches}

Contemporary models have made notable progress in addressing these challenges. MagicNet \cite{MagicNet} incorporates causal graph networks with memory-aware components to capture spatiotemporal dependencies. CMIN \cite{cmin} leverages textual data and inter-stock causal relationships through innovative fusion techniques. HATR-I \cite{HATR-I} implements a sophisticated dual-attention mechanism combined with Hawkes processes and domain-aware graph modules. DeepClue \cite{deepclue} enhances model interpretability by visualizing influential prediction factors derived from news and social media data. Despite these advancements, persistent limitations remain in terms of: (1) scalability to large-scale financial datasets, (2) effective integration of heterogeneous data modalities, and (3) robustness to market anomalies and regime changes.

\subsection{Research Motivation}

Building upon these developments, we propose SPH-Net, which synergistically combines the hierarchical feature extraction capabilities of Vision Transformers (ViT) \cite{vit} with the sequential modeling strengths of Transformer architectures \cite{att}. Our framework introduces a novel co-attention mechanism operating across both temporal and feature dimensions, addressing key limitations in current approaches. Similar to established encoder-decoder frameworks \cite{aensi,sva,cnn-lstm,GRU-seq2seq}, SPH-Net employs ViT as the encoder for high-dimensional feature extraction, while utilizing a Transformer-based decoder for sequential prediction generation. This integrated approach aims to achieve superior predictive accuracy while maintaining robust generalization across diverse market conditions.

\section{Methodology}

As illustrated in Fig.~\ref{stru}, the SPH-Net architecture comprises two principal components: the Trend Recognition Feature Extractor and the Deep Time-Series Predictor. The input data undergoes initial preprocessing to transform it into a pixel matrix compatible with Vision Transformer (ViT) input requirements \cite{vit}. Subsequently, the processed data flows through a Transformer-based architecture \cite{att}, where the outputs can be decoded into various representations including graphical visualizations, numerical outputs, or other formats as required by specific application scenarios.

\subsection{Trend Recognition Feature Extractor}
While ViT \cite{vit} was originally designed for visual data processing, we adapt this architecture for time-series analysis by treating N-day stock price sequences as image-like patches. The transformation begins by segmenting the input time-series data $X\in\mathbb{R}^{T \times d}$ into $p$ distinct patches, as formalized in Eq.~\ref{xrd}:
\begin{equation}
    X = [X_1, X_2, ..., X_P], \quad X_i \in \mathbb{R}^{p \times d}
    \label{xrd}\\   
\end{equation} 
where $T$ denotes the temporal dimension and $d$ represents the feature dimensionality. The segmented data is then projected into a higher-dimensional space through the embedding operation defined in Eq.~\ref{embed}:
\begin{equation}
        Z = W_{E} X + b_{E}
    \label{embed}
\end{equation}
Here, $W_{E}$ constitutes a learnable projection matrix, $b_{E}$ represents the bias term, and $Z$ denotes the resulting embeddings for ViT processing. To compensate for the Transformer's inherent lack of sequential awareness, we incorporate learnable positional encodings as specified in Eq.~\ref{PE}:
\begin{equation}
    Z = Z + PE
    \label{PE}
\end{equation}
The positional encoding at time step $t$ is computed as:
\begin{equation}
    PE_{t} = W_{PE} t
    \label{pet}
\end{equation}

\subsection{Deep Time-Series Predictor}

The feature representations $H_{\text{ViT}}$, equivalently denoted as $ViT(Z)$, obtained from the ViT processing \cite{vit} are subsequently fed into a Transformer encoder \cite{att} to capture global temporal dependencies. The attention mechanism operates as follows:
\begin{equation}
    \text{Attention}(Q,K,V) = \text{softmax}\left(\frac{QK^{T}}{\sqrt{d_{k}}}\right)V
    \label{att-f}
\end{equation}
where $Q = H_{\text{ViT}} W_{Q}$, $K = H_{\text{ViT}} W_{K}$, and $V = H_{\text{ViT}} W_{V}$ represent the query, key, and value matrices respectively. The projection matrices $W_{Q}$, $W_{K}$, and $W_{V}$ are trainable parameters, while $d_{k}$ denotes the dimensionality of each attention head. The softmax operation computes attention weights, effectively determining the relative importance of different temporal segments.
The architecture further incorporates a multi-head self-attention mechanism (MHSA) to capture diverse temporal patterns:
\begin{align}
    \text{MHSA}(H_{\text{ViT}}) = \text{Concat}(\text{head}_{1}, ..., \text{head}_{h}) W_{O} \label{mhsa}\\
    \text{head}_{i} = \text{Attention}(H_{\text{ViT}} W_{Q}^{i}, H_{\text{ViT}} W_{K}^{i}, H_{\text{ViT}} W_{V}^{i}) \label{head}
\end{align}
To enhance training stability and mitigate gradient-related issues, the Transformer employs layer normalization (LayerNorm) followed by a feedforward network (FFN):
\begin{align}
H_{\text{ViT}}' = \text{LayerNorm}(H_{\text{ViT}} + \text{MHSA}(H_{\text{ViT}})) \\
\text{FFN}(H_{\text{ViT}}') = \sigma(W_{1}H_{\text{ViT}}' + b_{1})W_{2} + b_{2} \\
H_{\text{T}} = \text{LayerNorm}(H_{\text{ViT}}' + \text{FFN}(H_{\text{ViT}}'))
\end{align}
The final temporal features $H_{\text{T}}$ are subsequently transformed into stock price predictions through a fully connected layer:
\begin{equation}
    \hat{y} = W_{y} H_{\text{T}} + b_{y}
    \label{pre}
\end{equation}
where $W_{y}$ and $b_{y}$ represent the output layer's weight matrix and bias vector, respectively.

\begin{figure*}[ht]
  \centering
  \begin{subfigure}{0.45\linewidth}
    \centerline{\includegraphics[width=\columnwidth]{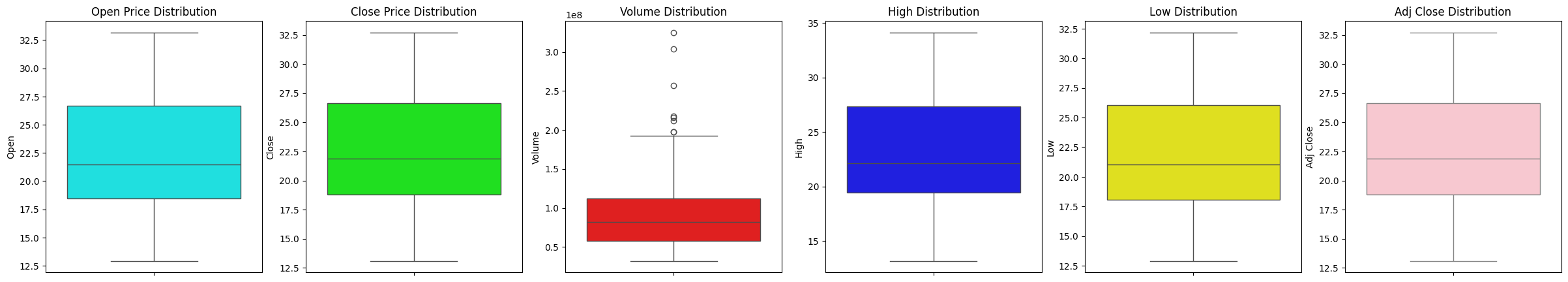}}
    \caption{Stock Price Distribution of AMD}
    \label{amd-1}
  \end{subfigure}
  \hfill
  \begin{subfigure}{0.45\linewidth}
    \centerline{\includegraphics[width=\columnwidth]{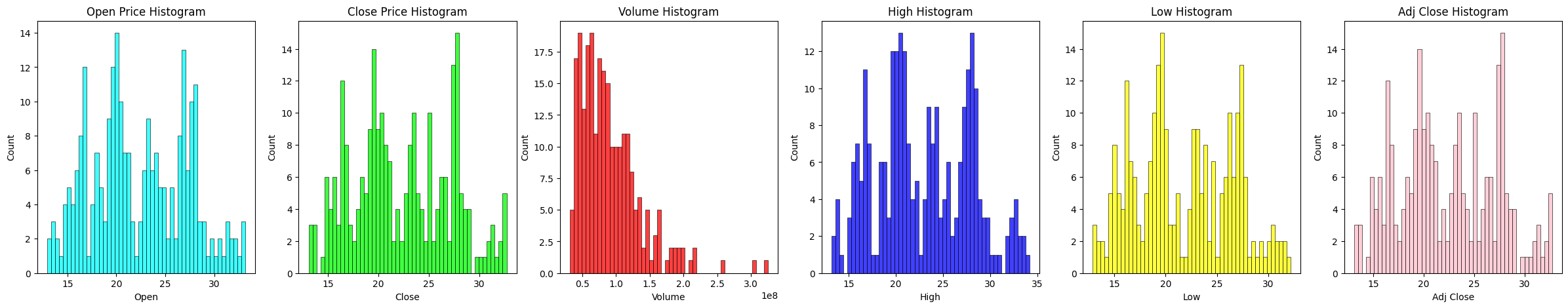}}
    \caption{Stock Price Distributions (Histogram) of AMD}
    \label{amd-2}
  \end{subfigure}  
  \hfill
  \begin{subfigure}{0.45\linewidth}
    \centerline{\includegraphics[width=\columnwidth]{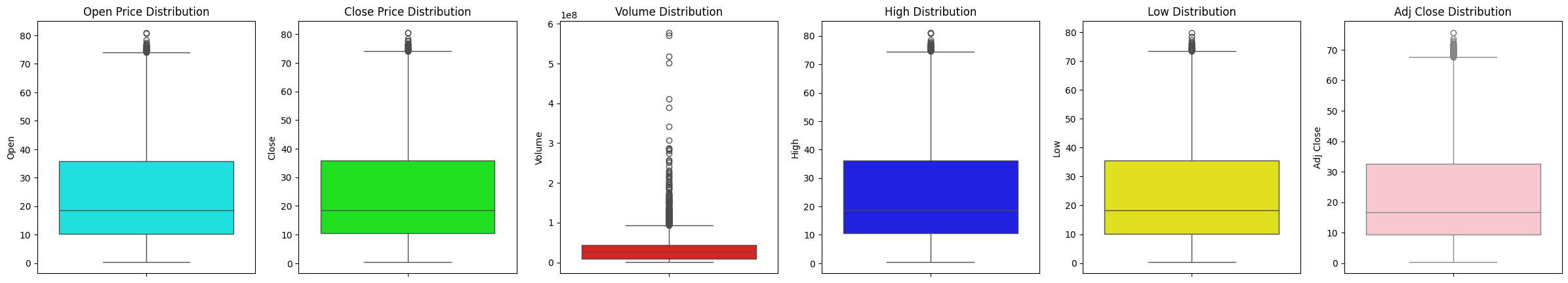}}
    \caption{Stock Price Distribution of Ebay}
    \label{ebay-1}
  \end{subfigure}
  \hfill
  \begin{subfigure}{0.45\linewidth}
    \centerline{\includegraphics[width=\columnwidth]{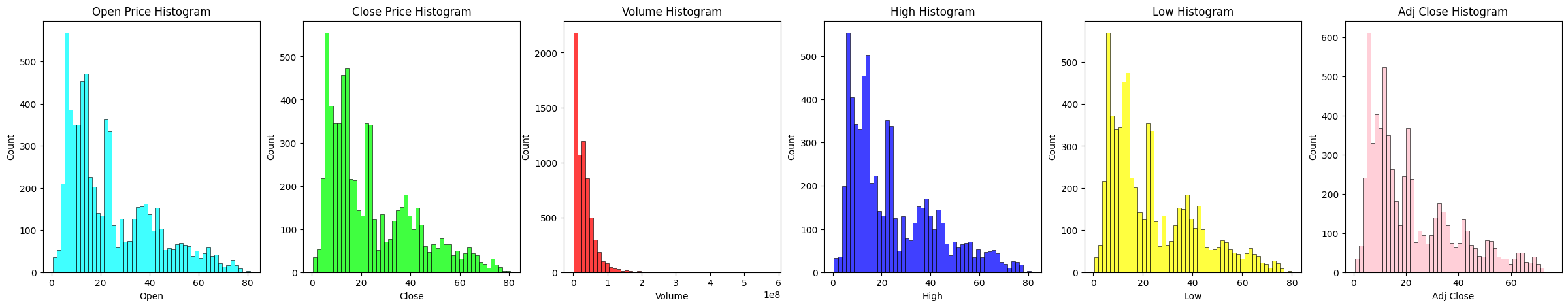}}
    \caption{Stock Price Distributions (Histogram) of Ebay}
    \label{ebay-2}
  \end{subfigure}
    \hfill
  \begin{subfigure}{0.45\linewidth}
    \centerline{\includegraphics[width=\columnwidth]{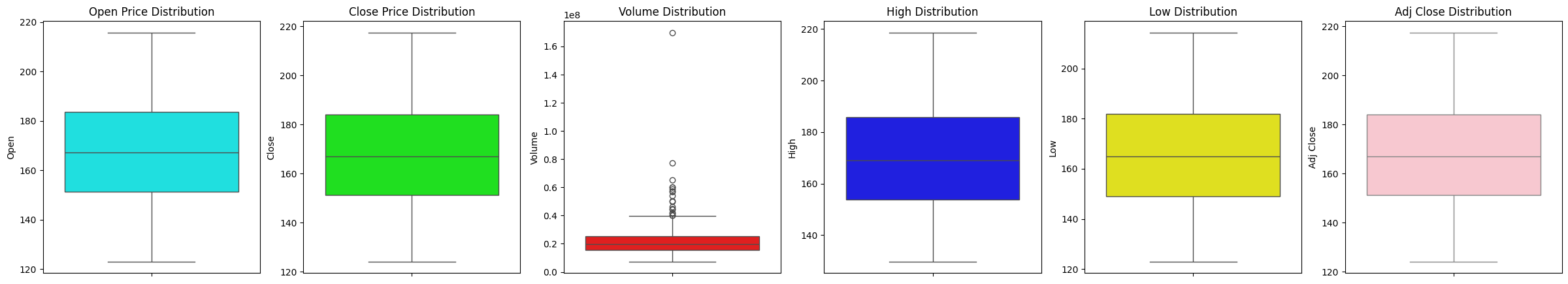}}
    \caption{Stock Price Distribution of FB}
    \label{fb-1}
  \end{subfigure}
  \hfill
  \begin{subfigure}{0.45\linewidth}
    \centerline{\includegraphics[width=\columnwidth]{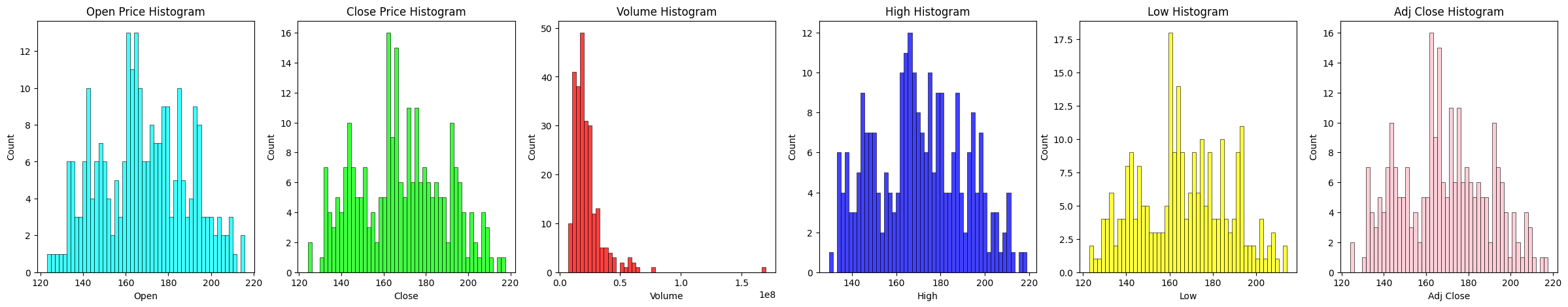}}
    \caption{Stock Price Distributions (Histogram) of FB}
    \label{fb-2}
  \end{subfigure}
      \hfill
  \begin{subfigure}{0.45\linewidth}
    \centerline{\includegraphics[width=\columnwidth]{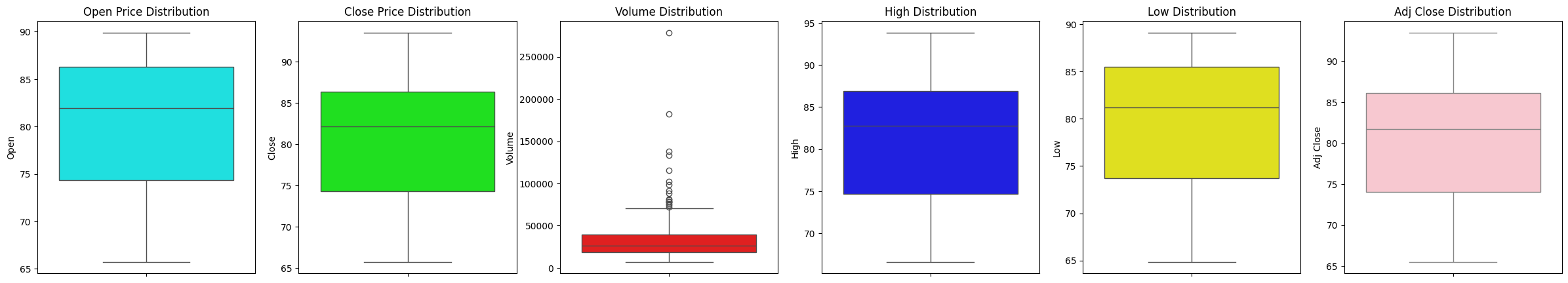}}
    \caption{Stock Price Distribution of FSV}
    \label{fsv-1}
  \end{subfigure}
  \hfill
  \begin{subfigure}{0.45\linewidth}
    \centerline{\includegraphics[width=\columnwidth]{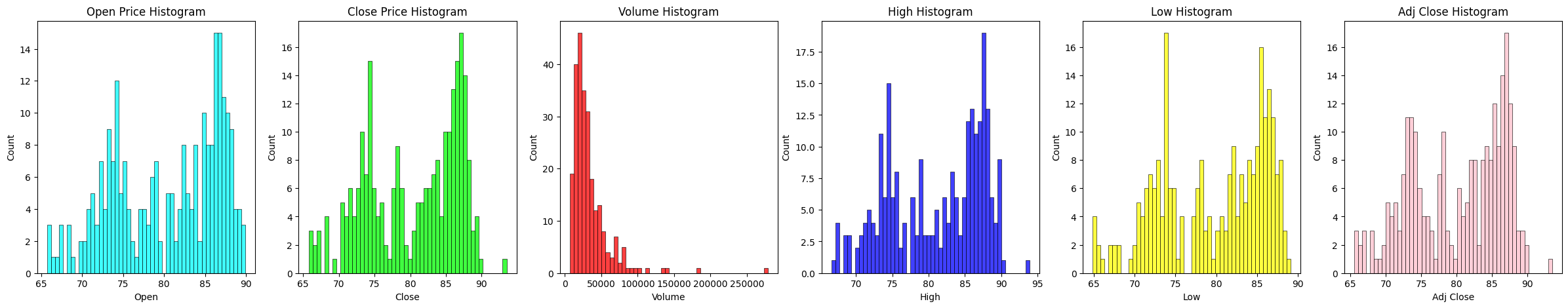}}
    \caption{Stock Price Distributions (Histogram) of FSV}
    \label{fsv-2}
  \end{subfigure}
        \hfill
  \begin{subfigure}{0.45\linewidth}
    \centerline{\includegraphics[width=\columnwidth]{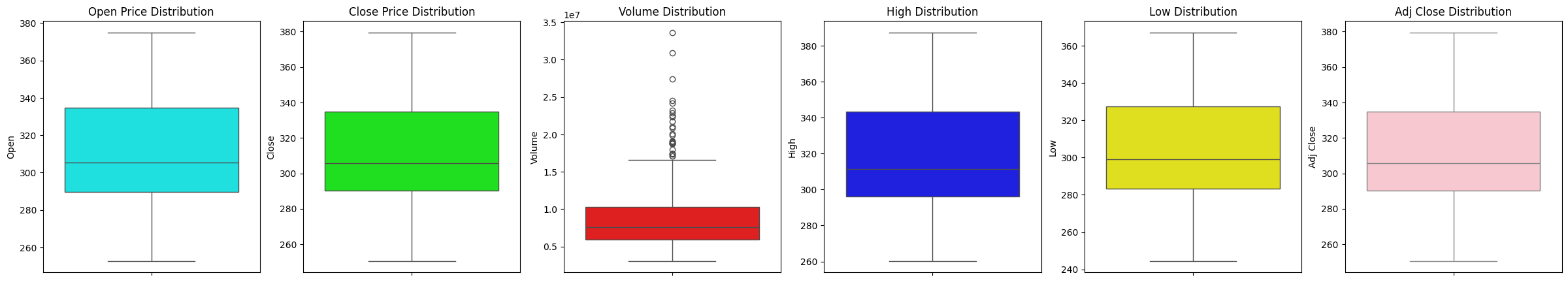}}
    \caption{Stock Price Distribution of Tesla}
    \label{tesla-1}
  \end{subfigure}
  \hfill
  \begin{subfigure}{0.45\linewidth}
    \centerline{\includegraphics[width=\columnwidth]{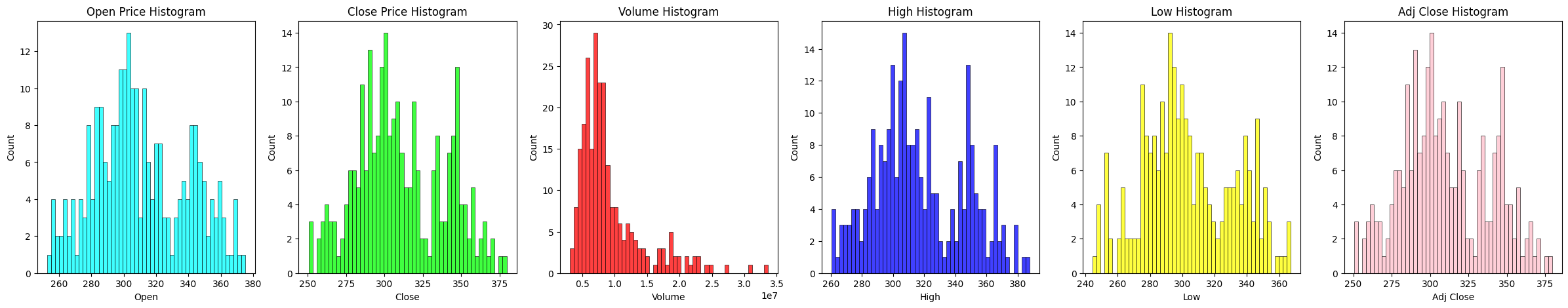}}
    \caption{Stock Price Distributions (Histogram) of Tesla}
    \label{tesla-2}
  \end{subfigure}
        \hfill
  \begin{subfigure}{0.45\linewidth}
    \centerline{\includegraphics[width=\columnwidth]{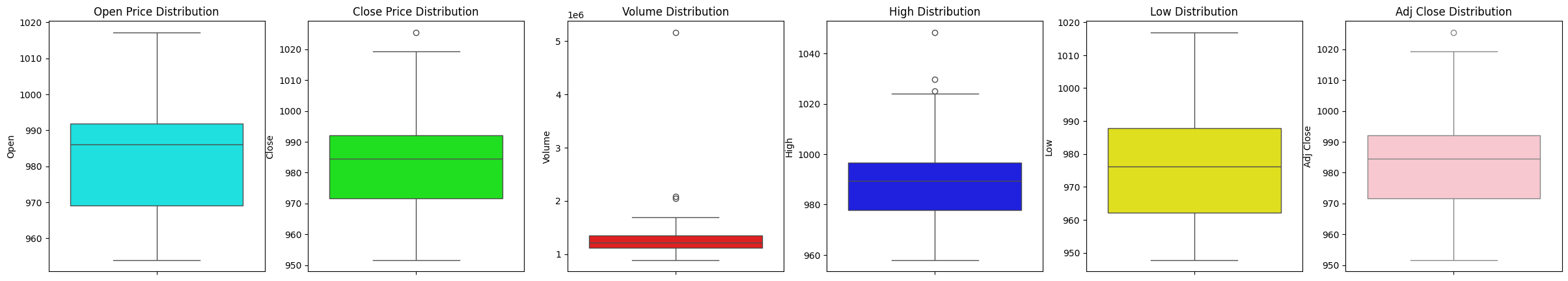}}
    \caption{Stock Price Distribution of Google}
    \label{gg-1}
  \end{subfigure}
  \hfill
  \begin{subfigure}{0.45\linewidth}
    \centerline{\includegraphics[width=\columnwidth]{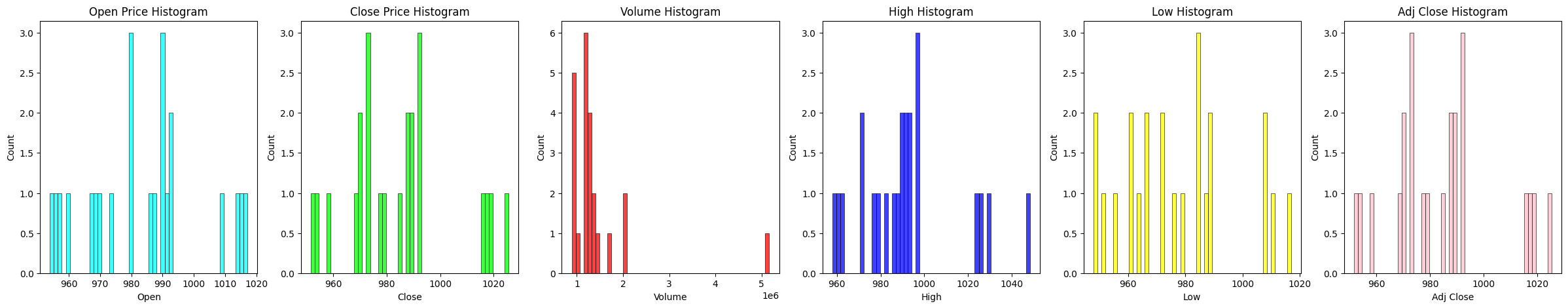}}
    \caption{Stock Price Distributions (Histogram) of Google}
    \label{gg-2}
  \end{subfigure}
        \hfill
  \begin{subfigure}{0.45\linewidth}
    \centerline{\includegraphics[width=\columnwidth]{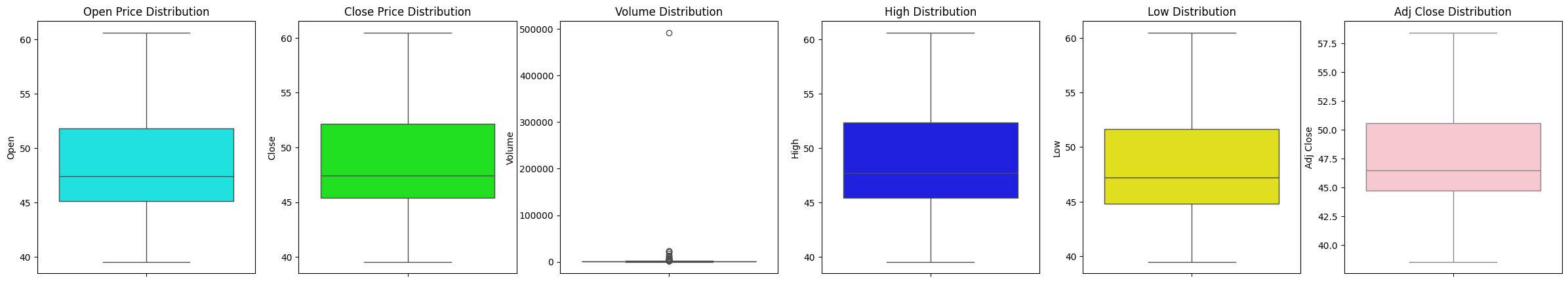}}
    \caption{Stock Price Distribution of Mondy}
    \label{mondy-1}
  \end{subfigure}
  \hfill
  \begin{subfigure}{0.45\linewidth}
    \centerline{\includegraphics[width=\columnwidth]{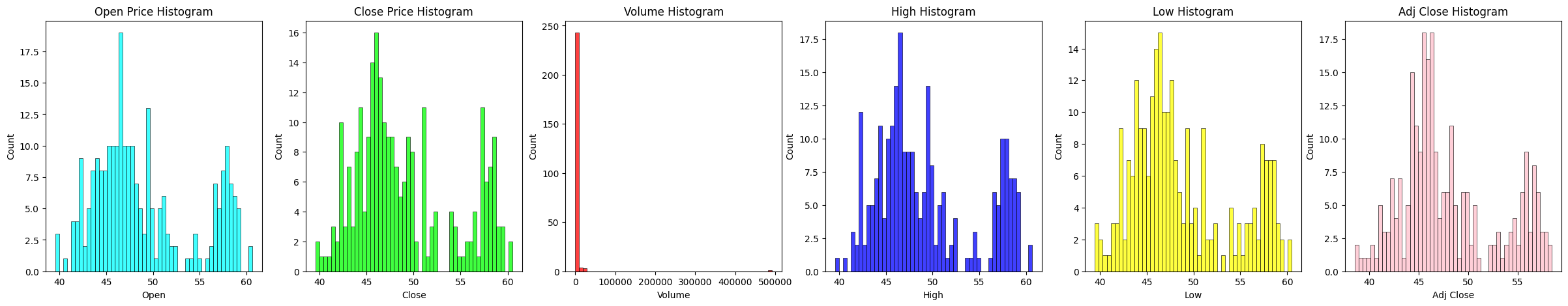}}
    \caption{Stock Price Distributions (Histogram) of Mondy}
    \label{mondy-2}
  \end{subfigure}
          \hfill
  \begin{subfigure}{0.45\linewidth}
    \centerline{\includegraphics[width=\columnwidth]{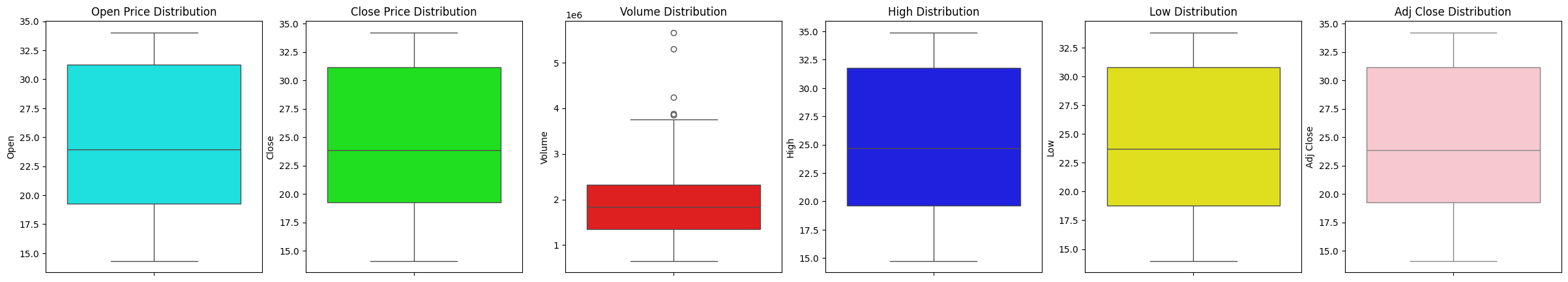}}
    \caption{Stock Price Distribution of MTDR}
    \label{mtdr-1}
  \end{subfigure}
  \hfill
  \begin{subfigure}{0.45\linewidth}
    \centerline{\includegraphics[width=\columnwidth]{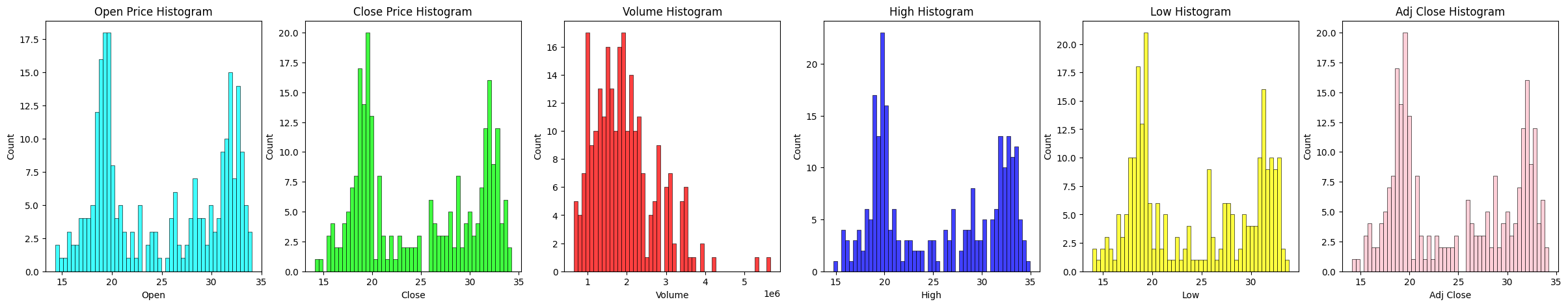}}
    \caption{Stock Price Distributions (Histogram) of MTDR}
    \label{mtdr-2}
  \end{subfigure}
\caption{Statistical distribution of stock prices across eight datasets.} 
  \label{sdsp}
\end{figure*}

\section{Dataset and Implementation}
\subsection{Dataset}

The experimental data is sourced from \textit{the Center for Research in SecurityPrices, LLC}\footnote{\url{https://www.crsp.org/}} (CRSP). Our investigation encompasses stock market data from eight publicly traded companies spanning diverse industrial sectors, including semiconductor technology, e-commerce platforms, social media, real estate investment trusts, electric vehicle manufacturing, internet services, sustainable packaging, and energy exploration. The selected companies (AMD, Ebay, FB, FSV, Tesla, Google, Mondy, and MTDR) provide comprehensive coverage of major market sectors, ensuring robust evaluation of our model across varying market conditions and industrial landscapes. 

The dataset comprises six standardized financial metrics: \textit{Open}, \textit{High}, \textit{Low}, \textit{Close}, \textit{Adj Close}, and \textit{Volume}. These metrics collectively provide a multidimensional view of market dynamics: the opening and closing prices bookend daily trading activity, while the high and low values delineate the intraday price range. The adjusted close price incorporates corporate actions such as dividends and stock splits, facilitating accurate historical comparisons. Trading volume serves as a crucial indicator of market liquidity and investor interest. These metrics are intrinsically interrelated - for instance, the high-low differential quantifies daily volatility, while the open-close relationship indicates prevailing market sentiment.

Fig.~\ref{sdsp} presents the statistical distributions of stock prices across all eight datasets, visualized through both price range charts and histograms. These visualizations reveal key characteristics of each stock's price behavior, including central tendency, dispersion, and outlier patterns. The comprehensive statistical analysis of these financial metrics forms the foundation for technical analysis and predictive modeling \cite{mxl,my}.

\subsection{Implementation}

\subsubsection{Data Preprocessing}
The implementation pipeline begins with data ingestion from eight CSV files, from which we extract the six key financial features. Missing values are addressed through forward-filling imputation, followed by MinMax normalization\footnote{\url{https://scikit-learn.org/stable/modules/generated/sklearn.preprocessing.MinMaxScaler.html}} to ensure numerical stability during training. We employ a sliding window approach to convert the time-series data into sequential samples, which are then transformed into PyTorch\footnote{\url{https://pytorch.org/}} tensors. The Dataset and DataLoader utilities facilitate efficient batch processing during model training.

\subsubsection{Model Configuration}
The model architecture employs a patch embedding dimension of 128, with the Transformer component consisting of 4 layers and 8 parallel attention heads. The hidden dimension (d\_model) is set to 128 to effectively capture complex temporal dependencies. Optimization is performed using the Adam algorithm \cite{adam} with a learning rate of 0.001 and a batch size of 32.

\subsubsection{Evaluation Matrix}

We employ two principal metrics for model evaluation:

\noindent\textbf{Mean Squared Error (MSE)}: This metric quantifies the average squared deviation between predicted and actual stock prices, as defined in Eq.~\ref{mse}:

\begin{equation}
    MSE = \frac{1}{n} \sum_{i=1}^{n} (y_i - \hat{y}_i)^2
   \label{mse}
\end{equation}

where $y_{i}$ denotes the observed stock price, $\hat{y}_i$ represents the model's prediction, and $n$ indicates the number of observations. The squaring operation emphasizes larger errors, making MSE particularly sensitive to significant prediction inaccuracies.

\noindent\textbf{Coefficient of Determination (R$^{2}$)}: This metric evaluates the proportion of variance in stock prices explained by the model, as formalized in Eq.~\ref{r2}:

\begin{equation}
    R^2 = 1 - \frac{SS_{res}}{SS_{tot}}
   \label{r2}
\end{equation}

where $SS_{res}$ quantifies the residual variation unexplained by the model, and $SS_{tot}$ represents the total variation in the observed data. An R$^{2}$ value approaching 1 indicates excellent explanatory power, while values near 0 suggest limited predictive capability.

To evaluate the directional accuracy of our predictions (i.e., whether the model correctly predicts upward or downward price movements compared to actual market movements), we frame this assessment as a binary classification task. This approach allows us to measure the model's ability to correctly identify market trends using a confusion matrix, which provides a comprehensive performance evaluation through the following metrics:

\begin{itemize}
    \item \emph{True Positive (TP)}: the \emph{TP} indicates that Class \emph{A} is correctly identified as belonging to Class \emph{A}.
    \item \emph{True Negative (TN)}: this matrix indicates that Class \emph{A} is correctly identified as not belonging to Class \emph{A}.
    \item \emph{False Positive (FP)}: it indicates that Class \emph{A} is not correctly identified as belonging to Class \emph{A}.
    \item \emph{False Negative (FN)}: it indicates that Class \emph{A} is not correctly identified as not belonging to Class \emph{A}.
\end{itemize}

\begin{table*}[ht]
\begin{tiny}
\begin{center}
\caption{Comparative experiments of SPH-Net and other models on 8 datasets. ($\times$100\%)}
\label{er-1}
\begin{tabular}{c|cc|cc|cc|cc|cc|cc|cc|cc}
\toprule
\multirow{4.5}{*}{\textbf{Model}}  & \multicolumn{16}{c}{\textbf{Dataset}} \\
\cmidrule{2-17}  & \multicolumn{2}{c|}{\textbf{AMD}} & \multicolumn{2}{c|}{\textbf{Ebay}} & \multicolumn{2}{c|}{\textbf{FB}} & \multicolumn{2}{c|}{\textbf{FSV}} & \multicolumn{2}{c|}{\textbf{TESLA}} & \multicolumn{2}{c|}{\textbf{Google}} & \multicolumn{2}{c|}{\textbf{Mondy}} & \multicolumn{2}{c}{\textbf{MTDR}} \\
\cmidrule{2-17}  &  \textbf{R$^{2}$} & \textbf{MSE} & \textbf{R$^{2}$} & \textbf{MSE} & \textbf{R$^{2}$} & \textbf{MSE} & \textbf{R$^{2}$} & \textbf{MSE} & \textbf{R$^{2}$} & \textbf{MSE} & \textbf{R$^{2}$} & \textbf{MSE} & \textbf{R$^{2}$} & \textbf{MSE} & \textbf{R$^{2}$} & \textbf{MSE} \\
\midrule
Linear Regression \cite{lir} & 97.69 & 57.27 & \underline{99.88} & \underline{32.44} & \underline{98.48} & 628.3 & 98.22 & 63.67 & 91.06 & 8047 & -1.11 & 50323 & 91.35 & 301.2 & 97.83 & 66.22\\
Logistic Regression \cite{lor} & 97.12 & 47.89 & 98.85 & 59.74 & 97.99 & 445.9 & 98.34 & 52.17 & 78.64 & 15690 & -7.23 & 69873 & 94.87 & 259.3 & 97.89 & 45.78\\
RNN \cite{rnn} & 97.12 & 39.89 & 98.73 & 41.27 & 98.12 & 712.9 & 97.18 & 45.89 &89.36 & 15743 & -3.28& 78963 & 91.87 & 378.4 & 96.84 & 57.14\\
Bi-RNN \cite{bi-rnn} & 98.46 & 38.32 & 99.01 & 39.68 & 98.41 & 814.9 & 96.89 & 59.69 & 82.78 & 8763 & -17.25 & 98895 & 95.69 & 433.6 & 92.75 & 174.9\\
CNN-RNN \cite{cnn-rnn} & 98.51 & 85.69 & 99.14 & 42.19 &98.36 & 684.3 & 98.14 & 106.9 & 92.65 & 11065 & -3.16 & 49595  & 96.11 & 239.8 & 96.18 & 58.35\\
LSTM \cite{lstm} & 97.53 & 44.58 & 98.42 & 39.17 & 97.25 & 417.9 & 97.65& 101.3 & 93.62 & 7471 & -0.69 & 32956 & 95.58 & 271.6 & 95.84 & 87.48\\
Bi-LSTM \cite{bi-lstm} & 98.89 & 61.23 & 98.78 & 47.38 & 97.84 & 389.7 & 95.86 & 98.99 & 91.22 & 5549 & 0.85 & 24451 & 95.87 & 259.4 & 91.45&145.2\\
LSTM-seq2seq \cite{LSTM-seq2seq} & 96.26 & 59.63 & 96.85 & 51.25 & 95.97 & 409.6 & 87.41 & 102.1 & 87.99& 9541 & -0.29 & 33326 & 95.69 & 396.5 &92.84 & 163.5\\
CNN-LSTM \cite{cnn-lstm} & 98.69 & 36.95 & 98.79 & 44.57 & 98.14 & 563.4 & \underline{98.89} & 44.19 & 94.41 & 4502  & -0.03 & 29589 & 95.82 & 393.15 & 95.69 & 112.8\\
CMIN \cite{cmin} & 97.99& 138.96 & 98.67 & 48.96 & 98.03 & 487.3 & 95.57 & 68.89 & 94.02 & 6998 & -0.98 & 59687 & 96.12 & 175.9 & 94.38 & 98.74\\
BERT \cite{bert} & \underline{99.19} & 38.11 & 99.34 & 54.98 & 98.22 & \underline{337.4} & 97.88 & 32.68 & \underline{94.55} &\underline{4295} & \underline{0.98} & \underline{9995} & \underline{96.14}& 215.7 & \underline{99.09} & \underline{38.65}\\
GRU \cite{gru} & 97.86 & 45.62& 98.71 & 94.10& 96.89 & 786.8 & 94.35 & 69.59 & 78.26 & 56326 & -12.3 & 14036 & 96.02 & 211.8 & 96.85 & 69.38\\
GRU-seq2seq \cite{GRU-seq2seq} & 98.12 & 71.75 & 98.82 & 69.85 & 97.13 & 1107 & 95.62 & 45.29 & 93.87 & 6563 & -0.99& 15604 & 95.67 & 210.6 & 93.59 & 58.49\\
Transformer \cite{att} & 98.97 & 35.94 & 99.16 & 32.74 & 98.13 & 374.9 & 97.89 & 46.36 & 93.36 & 5886& -0.32 & 11020 & 94.89 & 185.9 & 98.49 & 51.86\\
DeepClue \cite{deepclue} & 98.76 & 36.10 & 98.73& 132.6& 97.79 & 518.9 & 98.88 & 42.62 & 94.25 & 5447 & 0.95& 13212 & 95.72 & 298.6  & 98.56 & 47.85\\
HATR-I \cite{HATR-I} & 98.74 & \underline{35.41} & 98.98 & 75.91 & 98.39 & 478.6 & 98.21 & \underline{42.12} & 93.85 &6859 & 0.23 & 11953 & 95.81 & \underline{132.5} & 98.74 & 41.85\\
MagicNet \cite{MagicNet} & 98.47 & 39.78 & 97.36 & 68.96 & 97.98 & 419.8 & 96.87 & 68.96 & 81.97 & 14980 & 0.12& 10549& 92.64 & 236.9 & 98.06 & 143.82\\
Decision Tree \cite{dt} & 91.94 & 197.9 & 99.75 & 71.88 & 95.90 & 1702 & 94.41 & 199.9 & 66.89 & 29829 & -292.8 & 93582 & 84.68 & 533.4 & 95.48 & 138.1\\
Random Forest \cite{rf} & 96.57 & 84.25 & 99.86 & 38.26 & 97.84 & 894.0 & 97.14 & 102.1 & 86.19 & 12440 & -41.44 & 33694 & 89.32 & 372.1 & 97.37 & 80.23\\
\midrule
SPH-Net & \textbf{99.49} & \textbf{31.23} & \textbf{99.91} & \textbf{25.69} & \textbf{99.47} & \textbf{323.6} & \textbf{99.01} & \textbf{31.26} & \textbf{94.59} & \textbf{4103} & \textbf{1.02}  & \textbf{9978} & \textbf{96.35} & \textbf{127.3} & \textbf{99.12} & \textbf{38.63}\\ 
\bottomrule
\end{tabular}
\end{center}
\end{tiny}
\end{table*}

\begin{table*}[h]
\begin{tiny}
\centering
\caption{Comparative experiments with other models for stock price prediction classification. ($\times$100\%)}
\begin{center}
\begin{tabular}{c|ccc|ccc|ccc|ccc}
\toprule
\multirow{3}{*}{\diagbox{\textbf{Dataset}}{\textbf{Model}}}& \multicolumn{3}{c|}{\textbf{SPH-Net}} & \multicolumn{3}{c|}{\textbf{DeepClue \cite{deepclue}}} & \multicolumn{3}{c|}{\textbf{MagicNet \cite{MagicNet}}} & \multicolumn{3}{c}{\textbf{HATR-I \cite{HATR-I}}}\\
\cmidrule{2-13}  & \textbf{Precision}& \textbf{Accuracy}& \textbf{Recall} & \textbf{Precision}& \textbf{Accuracy}& \textbf{Recall} & \textbf{Precision}& \textbf{Accuracy}& \textbf{Recall} & \textbf{Precision}& \textbf{Accuracy}& \textbf{Recall}\\
\midrule
\textbf{AMD} & \underline{96.64} & \textbf{94.84} & \textbf{94.21} & 94.81 & 93.40 & 88.79 & 91.57 & 87.26 & 84.98 & \textbf{97.32} & \underline{94.08} & \underline{93.04} \\
\textbf{Ebay} & \textbf{97.64} & \textbf{96.12} & \textbf{94.59} & \underline{97.35} & \underline{92.55} & \underline{89.25} & 93.21 & 90.42 & 88.25 & 89.27 & 85.58 & 81.57 \\
\textbf{FB} & \textbf{96.27} & \textbf{92.32} & \textbf{90.93} & \underline{91.43} & \underline{90.20} & \underline{87.00} & 90.74 & 86.19 & 83.79 & 86.52 & 84.95 & 80.90 \\
\textbf{FSV} & \textbf{98.26} & \textbf{96.11} & \textbf{94.56} & 94.11 & 92.82 & 88.53 & 94.18 & 92.86 & 91.52 & \underline{97.83} & \underline{95.33} & \underline{92.85} \\
\textbf{TESLA} & \textbf{97.17} & \textbf{96.12} & \textbf{93.21} & 94.79 & 92.29 & 90.95 & 95.10 & 91.87 & 89.17 & \underline{96.78} & \underline{95.34} & \underline{92.37} \\
\textbf{Google} & \textbf{95.03} & \textbf{92.09} & \textbf{88.12} & 86.54 & 85.07 & 81.48 & \underline{94.70} & \underline{91.37} & \underline{86.52} & 89.87 & 87.73 & 83.26 \\
\textbf{Mondy} & \textbf{98.15} & \textbf{97.09} & \underline{91.87} & \underline{97.61} & \underline{96.44} & \textbf{92.64} & 91.86 & 86.89 & 85.59 & 92.20 & 87.32 & 84.23 \\
\textbf{MTDR} & \textbf{97.23} & \textbf{94.11} & \textbf{91.31} & 93.16 & 89.82 & 85.22 & 85.59 & 83.47 & 78.67 & \underline{96.57} & \underline{93.75} & \underline{90.27} \\
\bottomrule
\end{tabular}
\label{er-1-c}
\end{center}
\end{tiny}
\end{table*}

Especially in classification tasks, three key evaluation metrics are precision, accuracy, and recall. For precision, it measures how many of the predicted positive cases are actually positive. High precision means fewer FP. Its formula is shown below in Eq.~\ref{pre}:
\begin{equation}
	\begin{split}
\label{pre}
\text{Precision} = \frac{TP}{TP + FP}
	\end{split}
\end{equation}
For accuracy, it measures the overall correctness of the model by calculating the percentage of correctly classified samples (both positive and negative). Its formula is shown below in Eq.~\ref{acc}:
\begin{equation}
	\begin{split}
\label{acc}
\text{Accuracy} = \frac{TP + TN}{TP + TN + FP + FN}
	\end{split}
\end{equation}
For recall, it measures how many actual positive cases were correctly identified by the model. Its formula is shown below in Eq.~\ref{rec}:
\begin{equation}
	\begin{split}
\label{rec}
\text{Recall} = \frac{TP}{TP + FN}
	\end{split}
\end{equation}

\subsubsection{Hardware}

the hardware specifications for training and testing include 4 4090-Ti GPUs (4 $\times$ 24GB), 64GB of RAM, 8 CPU cores per node, and a total of 6 nodes.

\begin{table*}[ht]
\begin{tiny}
\begin{center}
\caption{Ablation experiment of SPH-Net on 8 datasets. ($\times$100\%)}
\label{er-ab}
\begin{tabular}{c|cc|cc|cc|cc|cc|cc|cc|cc}
\toprule
\multirow{4.5}{*}{\textbf{Model}}  & \multicolumn{16}{c}{\textbf{Dataset}} \\
\cmidrule{2-17}  & \multicolumn{2}{c|}{\textbf{AMD}} & \multicolumn{2}{c|}{\textbf{Ebay}} & \multicolumn{2}{c|}{\textbf{FB}} & \multicolumn{2}{c|}{\textbf{FSV}} & \multicolumn{2}{c|}{\textbf{TESLA}} & \multicolumn{2}{c|}{\textbf{Google}} & \multicolumn{2}{c|}{\textbf{Mondy}} & \multicolumn{2}{c}{\textbf{MTDR}} \\
\cmidrule{2-17}  &  \textbf{R$^{2}$} & \textbf{MSE} & \textbf{R$^{2}$} & \textbf{MSE} & \textbf{R$^{2}$} & \textbf{MSE} & \textbf{R$^{2}$} & \textbf{MSE} & \textbf{R$^{2}$} & \textbf{MSE} & \textbf{R$^{2}$} & \textbf{MSE} & \textbf{R$^{2}$} & \textbf{MSE} & \textbf{R$^{2}$} & \textbf{MSE} \\
\midrule
ViT \cite{vit} & 98.23 & 39.67 & 98.65 & 48.36 & 97.35 & 402.5 & 96.25 & 52.39 & 90.78 & 6754 & -0.36 & 13642 & 93.14 & 244.7 & 97.82 & 59.22 \\
ViT (patches = 2) & 97.27 & 49.18 & 98.99 & 37.88 & 97.17 & 392.0 & 96.00 & 38.48 & 98.54 & 4132 & 0.49 & 19456 & 95.79 & 194.3 & 97.87 & 96.20\\
ViT (patches = 4) & 99.08 & 34.67 & 99.56 & 28.95 & 98.74 & 341.8 & 98.45& 36.59 & 94.15 & 5223 & 0.25 & 10364 & 95.03 & 152.9 & 97.86 & 45.74\\
ViT (patches = 16) & 98.97 & \underline{32.36} & 99.70 & 27.34 & 99.31 & 328.7 & 98.88 & 32.33 & 94.45 & 4565 & 0.98 & 16082 & 96.17 & 135.3 & 97.91 & 39.47\\
\midrule
ViT + RNN \cite{rnn} & 99.47 & 31.39 & 99.88 & 25.93 & \underline{99.45} & 324.4 & \underline{98.99} & \underline{31.41} & \underline{94.57} & 4169 & -0.01 & 29992 & \underline{96.32} & \underline{128.4} & 98.00 & 38.75 \\
ViT + GRU \cite{gru} & 98.71 & 34.43 & 100.34 & 29.57 & 97.75 & 330.6 & 96.97 & 34.20 & 94.54 & 5201 & 0.90 & 10409 & 96.23 & 143.2 & 97.92 & 40.04 \\
ViT + LSTM \cite{lstm} & 99.28 & 33.17 & 99.54 & 28.51 & 99.20 & 332.4 & 98.79 & 33.10 & 94.34 & 4895 & 0.95 & 10157 & 96.05 & 141.3 & 97.82 & 40.07 \\
ViT + seq2seq \cite{} & 99.33 & 32.69 & 99.63 & 27.80 & 99.27 & 330.2 & 98.85 & 32.64 & 94.40 & 4697 & 0.96 & 10112 & 96.12 & 137.9 & 97.98 & 39.71\\
ViT + Bi-RNN \cite{gru} & 98.21 & 35.05 & 99.27 & 29.69 & 96.41 & 325.6 & 95.26 & 34.70 & 95.02 & 5244 & 0.88 & 10604 & 96.17 & 141.5 & 96.28 & 39.53  \\
ViT + Bi-LSTM \cite{gru} & 99.35 & 32.52 & 99.66 & 27.57 & 99.29 & 329.5 & 98.87 & 32.48 & 94.43 & 4631 & 0.97 & \underline{10097} & 96.15 & 136.5 & 97.95 & 39.59 \\
\midrule
Transformer \cite{att} & 98.97 & 35.94 & 99.16 & 32.74 & 98.13 & 374.9 & 97.89 & 46.36 & 93.36 & 5886& -0.32 & 11020 & 94.89 & 185.9 & 98.01 & 51.86\\
Transformer (head=2) & 97.64 & 47.18 & 98.65 & 41.22 & 96.75 & 623.1 & 95.29 & 58.36 & 92.14 & 6345 & -1.52 & 12034 & 93.74 & 224.5 & 96.71 & 62.17 \\
Transformer (head=4) & 99.08 & 34.67 & 99.56 & 28.95 & 98.74 & 341.8 & 98.45 & 36.59 & 94.15 & 5223 & 0.25 & 10364 & 95.03 & 152.9 & 97.92 & 45.74 \\
Transformer (head=16) & 98.42 & 40.32 & 99.03 & 35.67 & 98.03 & 379.2 & 97.35 & 47.81 & 93.01 & 5987 & -0.55 & 11230 & 94.36 & 192.1 & 97.85 & 53.78 \\
\midrule
Bert \cite{bert} + Transformer & 98.85 & 33.11 & 98.58 & 27.68 & 97.92 & \underline{324.7} & 97.12 & 32.96 & 94.81 & 4669 & 0.95 & 10286 & 96.23 & 134.3 & 97.20 & 39.09 \\
CNN \cite{cnn-lstm} + Transformer & 97.15 & 48.03 & 99.57 & 41.38 & 95.42 & 614.9 & 93.61 & 59.21 & 92.51 & 6397 & -1.49 & 12260 & 94.99 & 221.2 & 95.48 & 61.38\\
Decision Tree \cite{dt} + Transformer & 98.47 & 36.59 & 98.87 & 32.87 & 96.78 & 369.4 & 96.16 & 47.04 & 93.74 & 5934 & -0.31 & 11227 & 96.15 & 183.76 & 97.24 & 51.20\\
Random Forest \cite{rf} + Transformer & \underline{99.45} & 37.55 & 99.85 & \underline{26.16} & 99.42 & 325.8 & 98.97 & 31.57 & 94.55 & 4235 & \underline{1.01} & 14007 & 96.30 & 129.5 & 98.02 & 58.87 \\
Linear Regression \cite{lir} + Transformer & 97.76 & 48.31 & 98.08 & 37.73 & 95.53 & 397.5 & 97.73 & 37.92 & 94.14 & \underline{4108} & 0.50 & 19098 & 98.48 & 197.2 & 99.82 & 97.43 \\
Logistic Regression \cite{lor} + Transformer & 99.31 & 32.85 & 99.60 & 28.04 & 99.25 & 330.8 & 98.83 & 32.79 & 94.38 & 4763 & 0.96 & 10127 & 96.10 & 138.4 & 97.93 & 39.83 \\
\midrule
ViT \cite{vit} + Transformer \cite{att}
 & 99.21 & 33.82 & 99.42 & 29.45 & 99.11 & 335.4 & 98.72 & 33.71 & 94.26 & 5159 & 0.92 & 10217 & 95.95 & 145.6 & \underline{98.08} & 40.55 \\
(patches = 2) + (head = 2) & 97.93 & 41.05 & \underline{99.89} & 35.81 & 96.68 & 373.9 & 95.63 & 48.51 & 93.39 & 6036 & -0.54 & 11441 & 95.61 & 189.8 & 96.60 & 53.10 \\
(patches = 2) + (head = 4) & 97.16 & 59.37 & 99.41 & 39.02 & 97.76 & 297.4 & 98.71 & 98.10 & 94.31 & 5982 & -0.09 & 14997 & 94.65 & 198.6 & 97.78 & 48.30 \\
(patches = 2) + (head = 8) & 99.26 & 33.33 & 99.51 & 28.75 & 99.18 & 333.1 & 98.77 & 33.25 & 94.32 & 4961 & 0.94 & 10172 & 96.03 & 142.7 & 97.66 & 40.19 \\
(patches = 2) + (head = 16) & 98.80 & 33.61 & 100.4 & 28.38 & 97.85 & 326.4 & 97.06 & 33.42 & 94.54 & 4869 & 0.54 & 17464 & 96.17 & 138.1 & 97.32 & 39.44  \\
(patches = 4) + (head = 2) & 99.30 & 33.01 & 99.57 & 28.27 & 99.22 & 331.1 & 98.81 & 32.94 & 94.36 & 4829 & 0.55 & 17142 & 96.08 & 139.8 & 98.07 & 39.95 \\
(patches = 4) + (head = 4) & 97.49 & 60.04 & 100.02 & 62.60 & 95.25 & 461.7 & 94.12 & 57.92 & 94.38 & 9034 & -1.83 & 29930 & 96.14 & 279.3 & 95.41 & 58.08  \\
(patches = 4) + (head = 8) & 98.47 & 57.94 & 98.19 & 52.14 & 97.93 & 474.8 & 97.53 & 56.26 & 94.32 & 8887 & -0.89 & 19021 & 94.12 & 285.9 & 97.89 & 59.57 \\
(patches = 4) + (head = 16) & 97.98 & 58.98 & 99.10 & 52.35 & 96.58 & 468.2 & 95.81 & 57.08 & 94.50 & 8960 & -0.87 & 19378 & 95.37 & 282.1 & 96.64 & 58.82 \\
(patches = 8) + (head = 2) & 97.49 & 60.04 & 100.02 & 52.56 & 95.25 & 462.5 & 94.12 & 57.92 & 95.08 & 9034 & -0.85 & 19742 & 96.64 & 279.5 & 95.41 & 58.08   \\
(patches = 8) + (head = 4) & 99.18 & 32.49 & 99.53 & 37.12 & 98.73 & 378.2 & 98.52 & 36.38 & 93.47 & 5499 & 0.28 & 19067 & 96.01 & 137.1 & 97.37 & \underline{39.15} \\
(patches = 8) + (head = 16) & \textbf{99.49} & \textbf{31.23} & \textbf{99.91} & \textbf{25.69} & \textbf{99.47} & \textbf{323.6} & \textbf{99.01} & \textbf{31.26} & \textbf{94.59} & \textbf{4103} & \textbf{1.02}  & \textbf{9978} & \textbf{96.35} & \textbf{127.3} & \textbf{98.12} & \textbf{38.63}\\
(patches = 16) + (head = 2) & 98.47 & 36.59 & 99.08 & 32.87 & 96.78 & 369.4 & 96.16 & 47.04 & 93.74 & 5934 & -0.31 & 11227 & 96.15 & 183.6 & 97.24 & 51.20 \\
(patches = 16) + (head = 4) & 99.04 & 39.72 & 97.82 & 36.39 & 98.42 & 355.8 & 97.96 & 51.72 & 94.53 & 5301 & 1.00 & 14022 & 95.27 & 130.7 & 97.24 & 51.99 \\
(patches = 16) + (head = 8) & 98.87 & 39.36 & 98.70 & 41.34 & 98.31 & 345.7 & 97.88 & 42.33 & 92.45 & 5565 & 0.11 & 17082 & 95.17 & 155.1 & 97.41 & 45.47 \\
(patches = 16) + (head = 16) & 97.94 & 59.38 & 98.83 & 48.50 & 97.39 & 973.9 & 97.15 & 59.17 & 94.50 & 9877 & -0.19 & 19942 & 96.08 & 197.5 & 97.45 & 47.46  \\
\midrule
SPH-Net & \textbf{99.49} & \textbf{31.23} & \textbf{99.91} & \textbf{25.69} & \textbf{99.47} & \textbf{323.6} & \textbf{99.01} & \textbf{31.26} & \textbf{94.59} & \textbf{4103} & \textbf{1.02}  & \textbf{9978} & \textbf{96.35} & \textbf{127.3} & \textbf{98.12} & \textbf{38.63}\\ 
\bottomrule
\end{tabular}
\end{center}
\end{tiny}
\end{table*}

\section{Experimental Result}

In this section, we conduct comprehensive experiments to evaluate the effectiveness and robustness of the proposed SPH-Net model. Our evaluation focuses on two key aspects: (1) regression performance for precise price prediction and (2) classification accuracy for trend direction identification. Through systematic comparisons with state-of-the-art baselines, detailed ablation studies, and analytical visualizations, we demonstrate the model's capability to capture complex market patterns.

All experiments follow a standardized protocol across eight stock datasets spanning multiple sectors (technology, energy, e-commerce, etc.). The datasets are partitioned into 70\% training and 30\% testing sets, with the model trained for 100 epochs using the Adam optimizer (learning rate = 0.001) and a batch size of 32. This consistent setup ensures fair comparison and reproducible results. 

\subsection{Comparative Experiment}

We first compare SPH-Net with traditional statistical methods (Linear Regression \cite{lir}, Logistic Regression \cite{lor}), machine learning baselines (Decision Tree \cite{dt}, Random Forest \cite{rf}), classical deep learning models (RNN \cite{rnn}, LSTM \cite{lstm}, GRU \cite{gru}, CNN-LSTM \cite{cnn-lstm}, Bi-LSTM \cite{bi-lstm}), and recent state-of-the-art models including Transformer \cite{att}, BERT \cite{bert}, DeepClue \cite{deepclue}, HATR-I \cite{HATR-I}, MagicNet \cite{MagicNet}, and CMIN \cite{cmin}.

Table~\ref{er-1} presents the regression results on eight datasets. SPH-Net consistently achieves the best R$^{2}$ values and lowest MSE scores across all cases. For instance, on the AMD dataset, SPH-Net achieves an R$^{2}$ of 0.9949 and MSE of 31.23, outperforming all baselines. On the Google dataset, SPH-Net is the only model with a positive R$^{2}$ (1.02), while most others perform poorly or even negatively due to high volatility and sparse data patterns. 

For the classification task of predicting price direction (up or down), we evaluate precision, accuracy, and recall (Table~\ref{er-1-c}). SPH-Net outperforms all competing models across all datasets. On TESLA, SPH-Net reaches a precision of 97.17\%, accuracy of 96.12\%, and recall of 93.21\%, substantially exceeding the closest competitor, DeepClue \cite{deepclue}. On FB, SPH-Net maintains a high classification precision of 96.27\%, demonstrating its robustness even on platforms heavily influenced by external events. These results clearly demonstrate SPH-Net's superior generalization ability, benefiting from its hybrid co-attention structure that effectively models both feature-wise and temporal dependencies.

\subsection{Ablation Experiment}

To investigate each architectural component's contribution, we conducted detailed ablation studies (Table~\ref{er-ab}). Following standard configurations, ViT's initial patches are set to 8 \cite{vit}, and the Transformer head number is also set to 8 \cite{att}. We examined:

\begin{itemize}
    \item \textbf{ViT patch sizes}: Testing sizes (2, 4, 8, 16) revealed optimal performance at 8, suggesting the best balance between context and granularity.
    \item \textbf{Transformer attention heads}: Evaluating configurations (2, 4, 8, 16) showed 8 heads provide the optimal trade-off between attention coverage and computational efficiency.
    \item \textbf{Alternative encoders}: Replacing the Transformer decoder with LSTM \cite{lstm}, GRU \cite{gru}, seq2seq \cite{seq2seq}, and BiLSTM \cite{bi-lstm} consistently resulted in inferior performance, validating the advantage of attention-based decoding.
\end{itemize}

The complete SPH-Net configuration, employing 8 ViT patches and 8 attention heads, achieves the best performance across all tasks.

\subsection{Visualization}

\begin{figure*}[ht]
  \centering
  \begin{subfigure}{0.20\linewidth}
    \centerline{\includegraphics[width=\columnwidth]{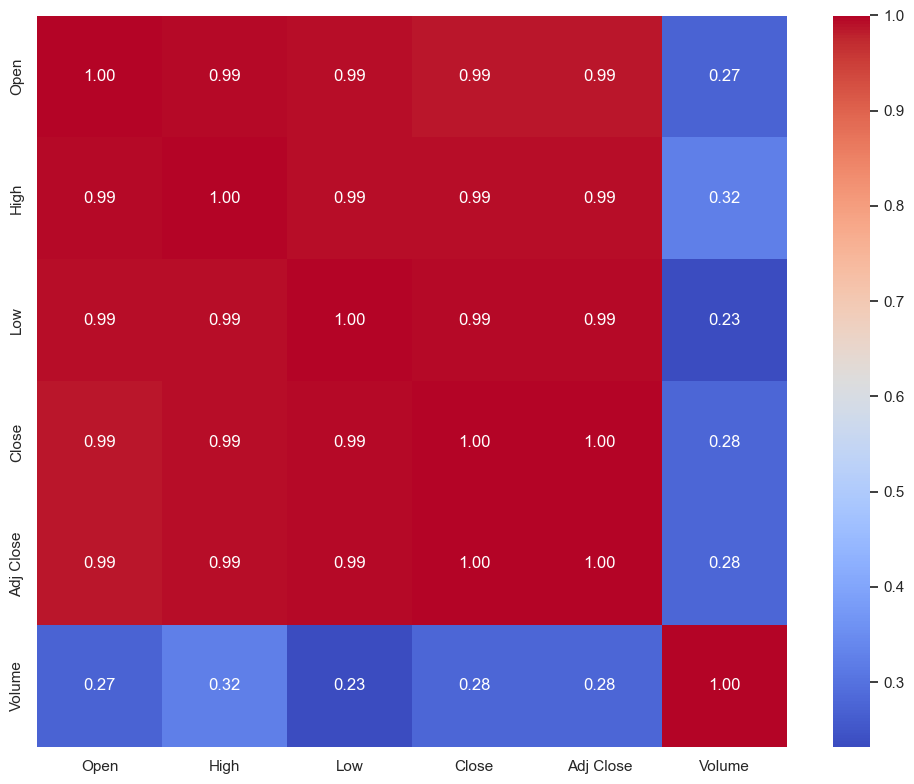}}
    \caption{Correlation Heatmap of AMD}
    \label{er-c-AMD}
  \end{subfigure}
  \hfill
  \begin{subfigure}{0.20\linewidth}
    \centerline{\includegraphics[width=\columnwidth]{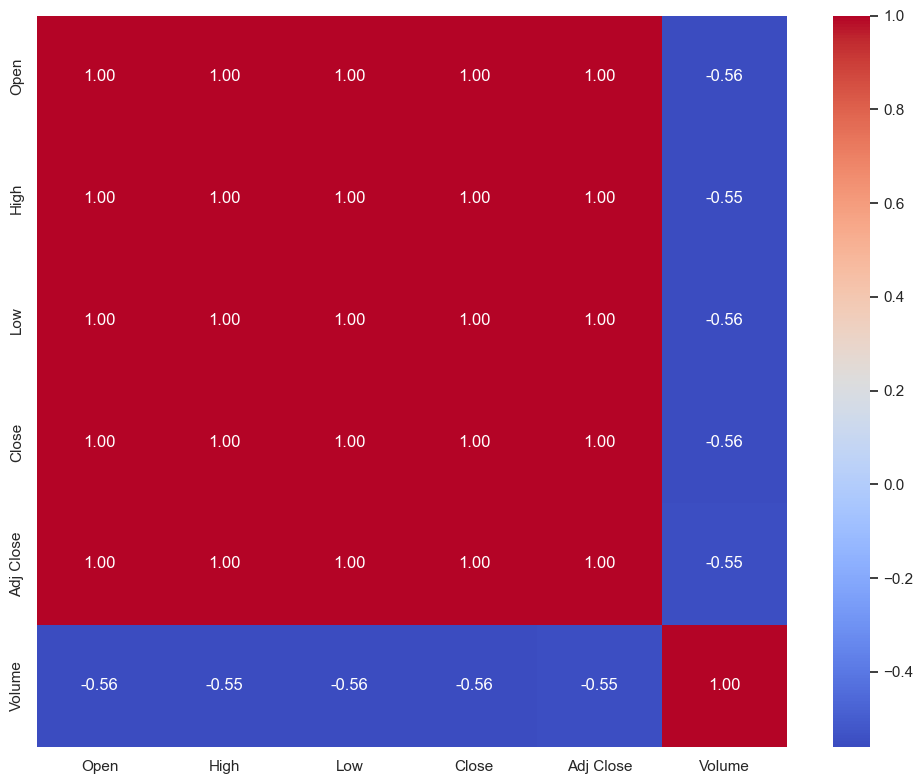}}
    \caption{Correlation Heatmap of Ebay}
    \label{er-c-Ebay}
  \end{subfigure}  
  \hfill
  \begin{subfigure}{0.20\linewidth}
    \centerline{\includegraphics[width=\columnwidth]{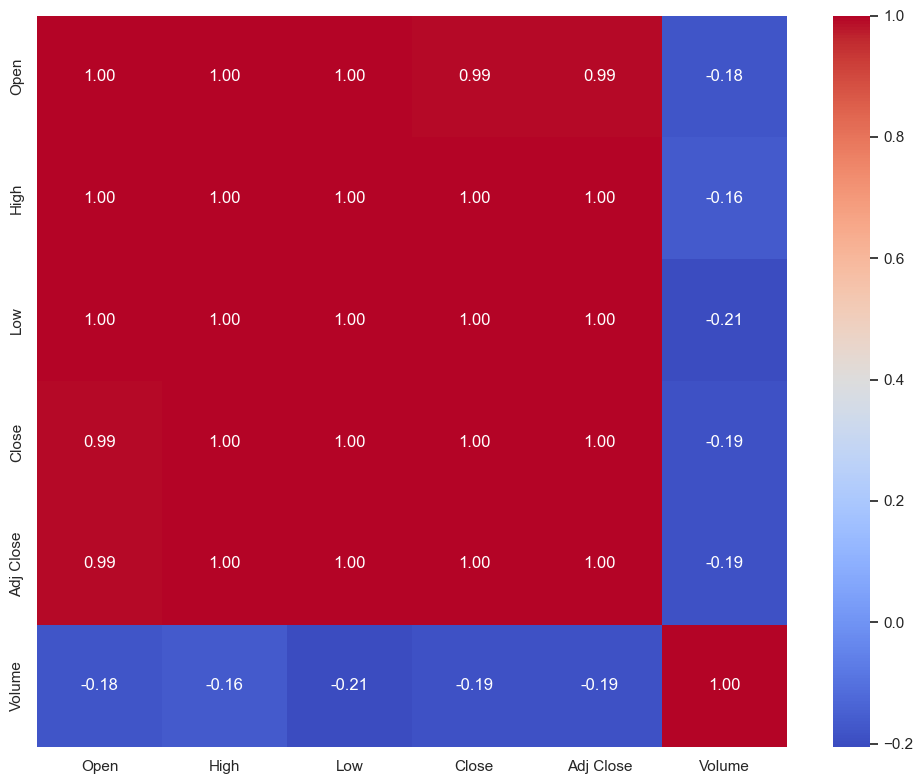}}
    \caption{Correlation Heatmap of FB }
    \label{er-c-FB}
  \end{subfigure}
  \hfill
  \begin{subfigure}{0.20\linewidth}
    \centerline{\includegraphics[width=\columnwidth]{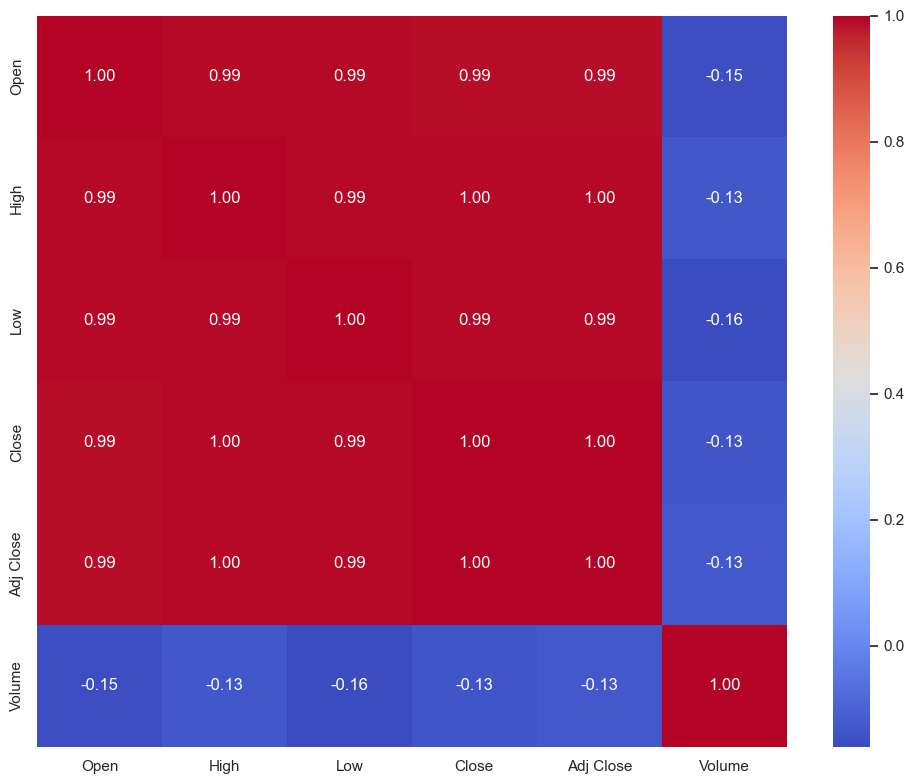}}
    \caption{Correlation Heatmap of FSV }
    \label{er-c-FSV}
  \end{subfigure}
    \hfill
  \begin{subfigure}{0.20\linewidth}
    \centerline{\includegraphics[width=\columnwidth]{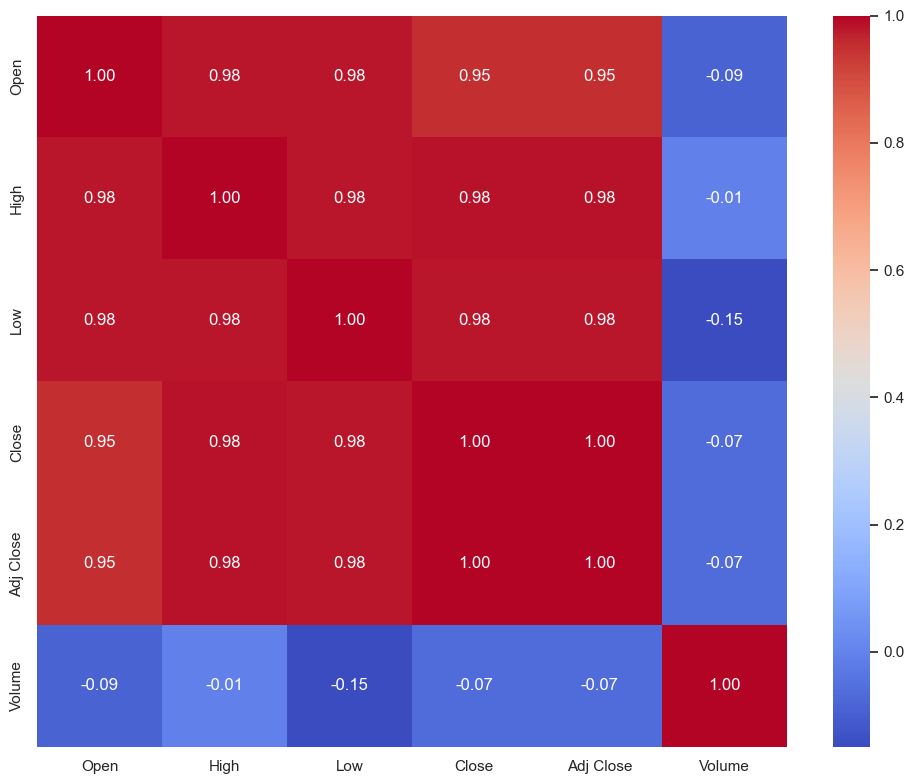}}
    \caption{Correlation Heatmap of Tesla}
    \label{er-c-tesla}
  \end{subfigure}
  \hfill
  \begin{subfigure}{0.20\linewidth}
    \centerline{\includegraphics[width=\columnwidth]{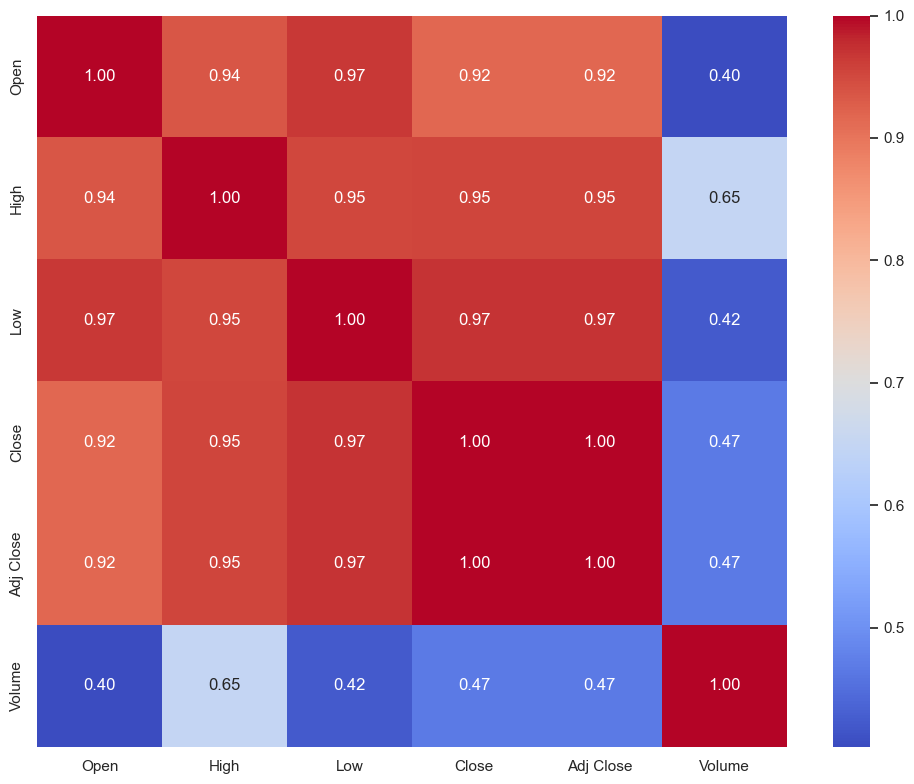}}
    \caption{Correlation Heatmap of Google}
    \label{er-c-Google}
  \end{subfigure}
      \hfill
  \begin{subfigure}{0.20\linewidth}
    \centerline{\includegraphics[width=\columnwidth]{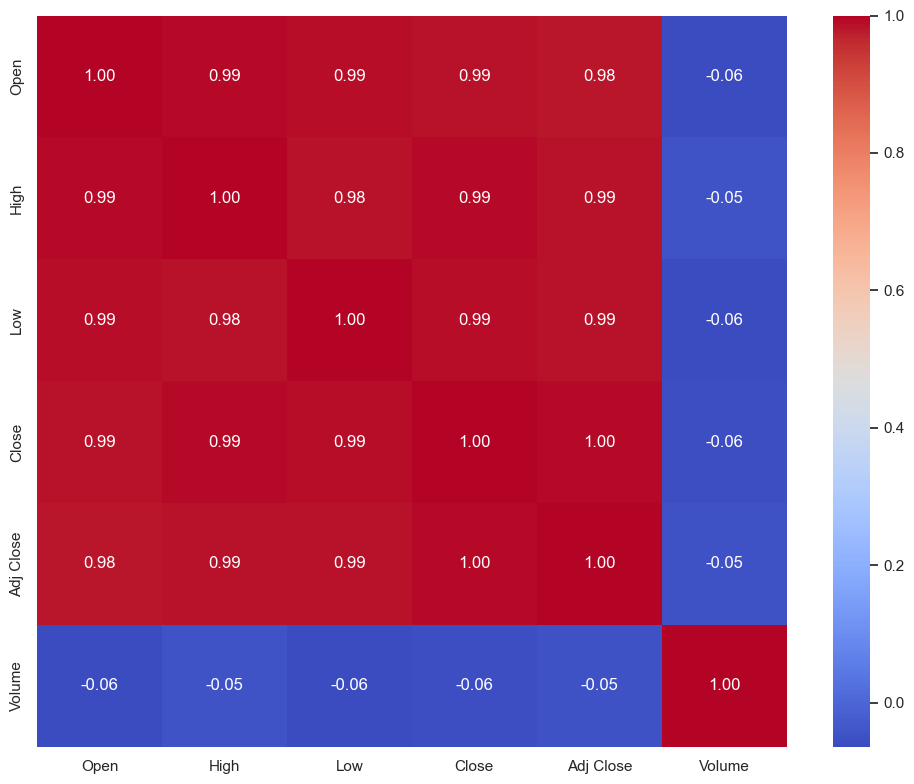}}
    \caption{Correlation Heatmap of Mondy}
    \label{er-c-Mondy}
  \end{subfigure}
  \hfill
  \begin{subfigure}{0.20\linewidth}
    \centerline{\includegraphics[width=\columnwidth]{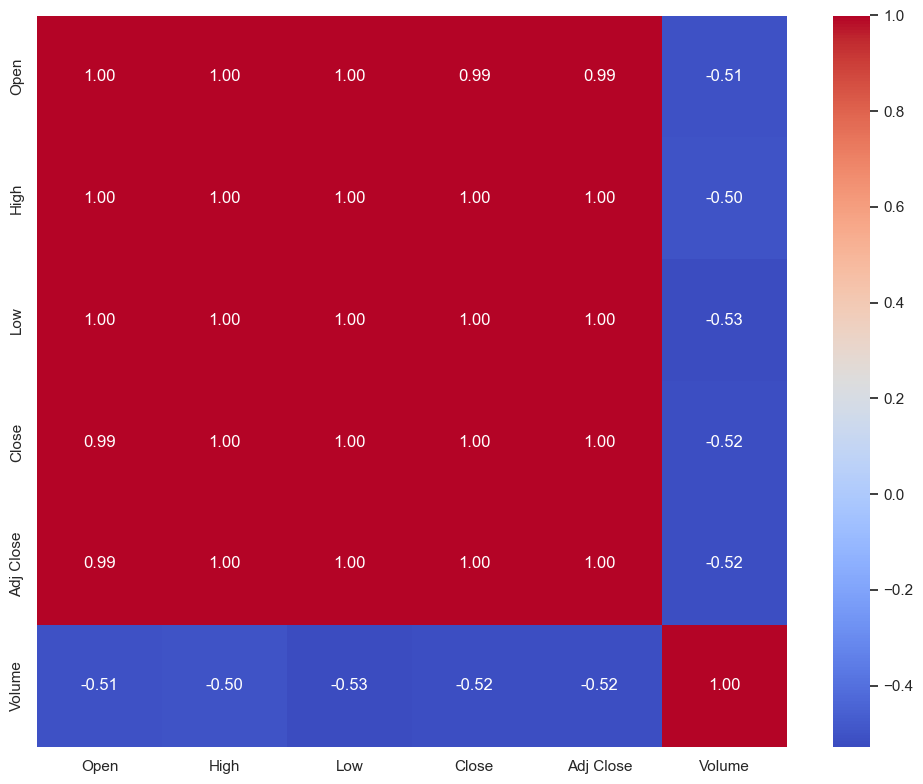}}
    \caption{Correlation Heatmap of MTDR}
    \label{er-c-MTDR}
  \end{subfigure}
\caption{Correlation heatmap of numerical features on 8 datasets.} 
  \label{er-c}
\end{figure*}

As shown in Fig.~\ref{er-c}, we generated correlation heatmaps for the six key numerical features on each dataset. These heatmaps provide insights into the linear relationships between features. For instance, Open and Close prices typically exhibit strong positive correlations, reflecting their co-movement throughout trading days. Meanwhile, Volume often shows weaker or variable correlations with price features, indicating its influence may be more context-dependent. These visualizations help identify which features are most informative for predicting future stock prices and assist in understanding feature dependencies.

\begin{figure*}[ht]
  \centering
  \begin{subfigure}{0.20\linewidth}
    \centerline{\includegraphics[width=\columnwidth]{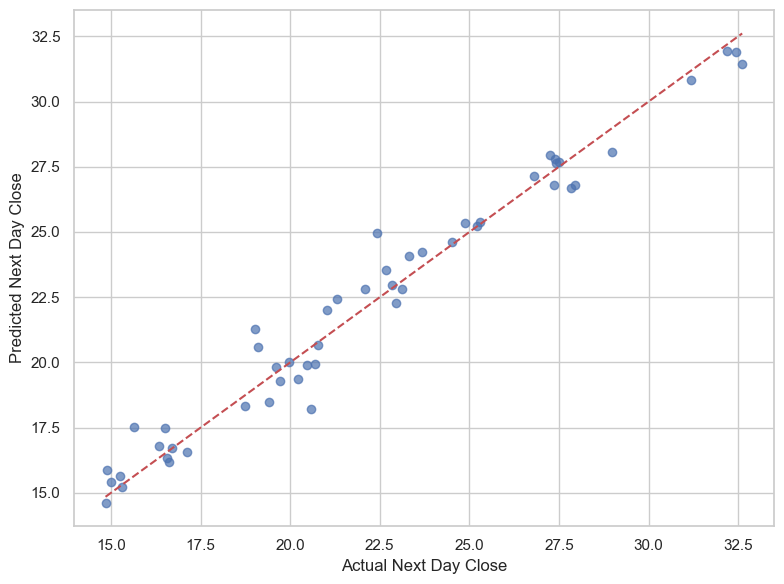}}
    \caption{Predicated V.S. Actual Prices on AMD}
    \label{er-np-AMD}
  \end{subfigure}
  \hfill
  \begin{subfigure}{0.20\linewidth}
    \centerline{\includegraphics[width=\columnwidth]{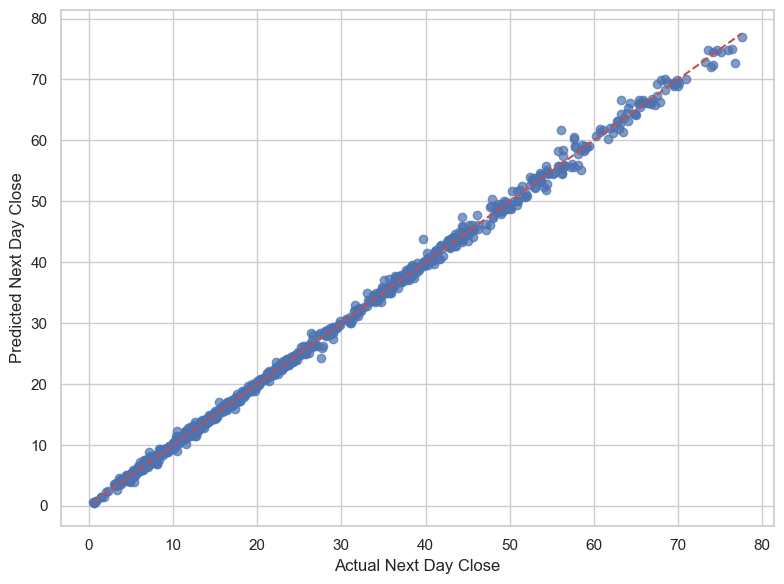}}
    \caption{Predicated V.S. Actual Prices on Ebay}
    \label{er-np-Ebay}
  \end{subfigure}  
  \hfill
  \begin{subfigure}{0.20\linewidth}
    \centerline{\includegraphics[width=\columnwidth]{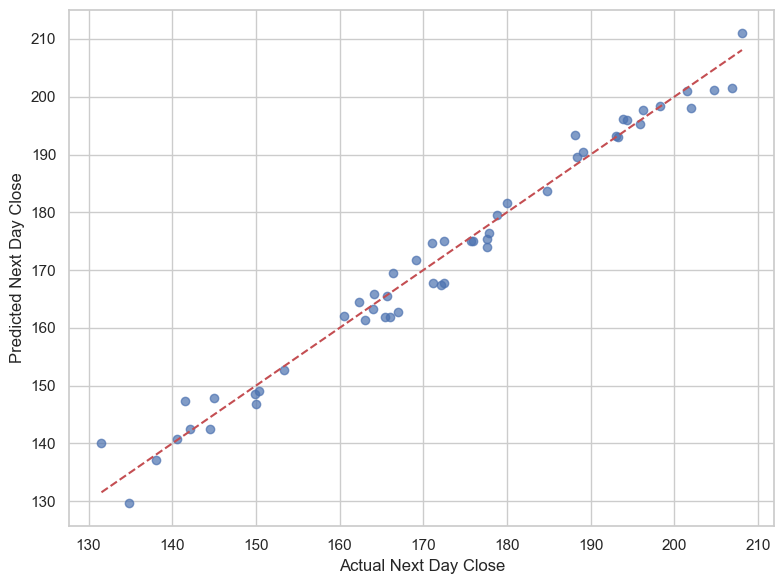}}
    \caption{Predicated V.S. Actual Prices on FB }
    \label{er-np-FB}
  \end{subfigure}
  \hfill
  \begin{subfigure}{0.20\linewidth}
    \centerline{\includegraphics[width=\columnwidth]{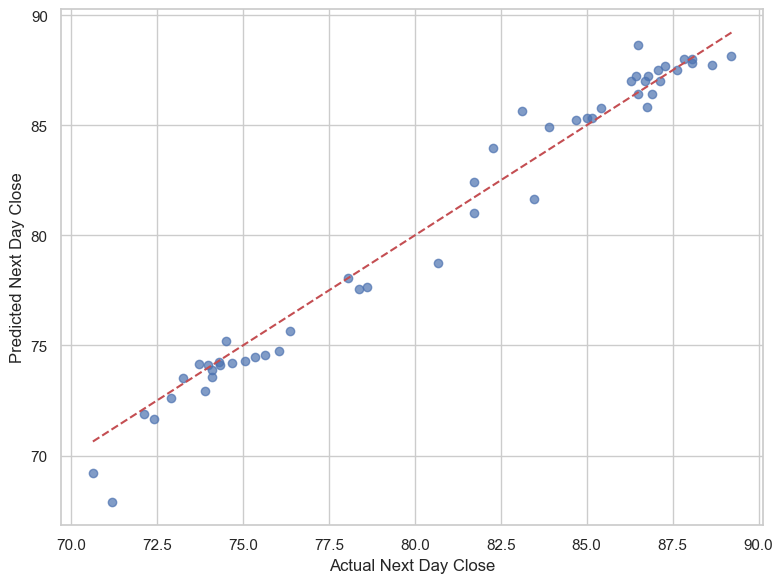}}
    \caption{Predicated V.S. Actual Prices on FSV }
    \label{er-np-FSV}
  \end{subfigure}
    \hfill
  \begin{subfigure}{0.20\linewidth}
    \centerline{\includegraphics[width=\columnwidth]{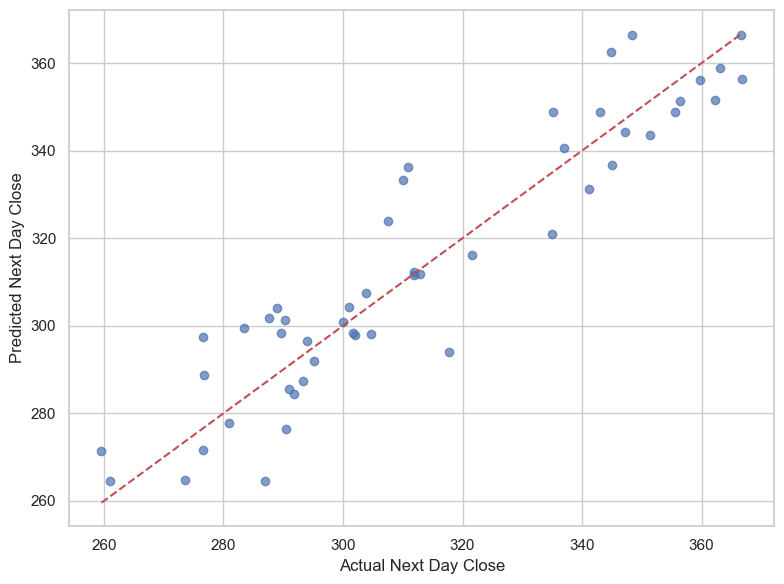}}
    \caption{Predicated V.S. Actual Prices on Tesla}
    \label{er-np-tesla}
  \end{subfigure}
  \hfill
  \begin{subfigure}{0.20\linewidth}
    \centerline{\includegraphics[width=\columnwidth]{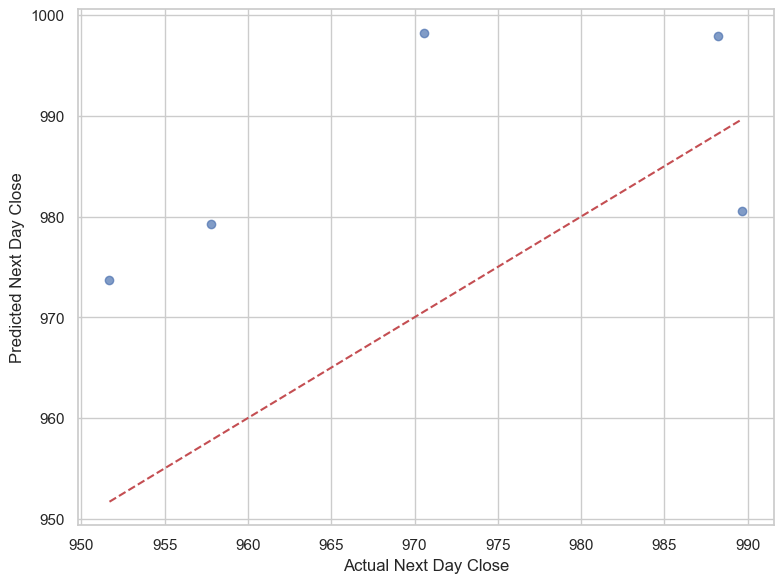}}
    \caption{Predicated V.S. Actual Prices on Google}
    \label{er-np-Google}
  \end{subfigure}
      \hfill
  \begin{subfigure}{0.20\linewidth}
    \centerline{\includegraphics[width=\columnwidth]{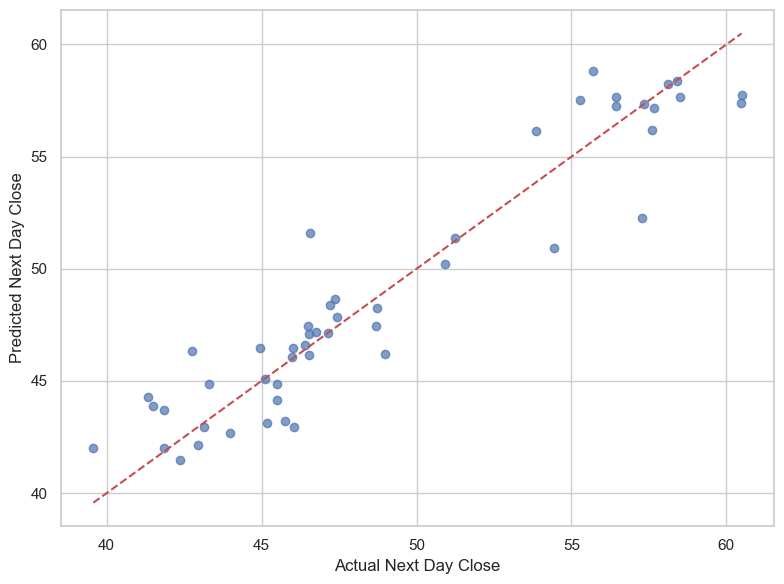}}
    \caption{Predicated V.S. Actual Prices on Mondy}
    \label{er-np-Mondy}
  \end{subfigure}
  \hfill
  \begin{subfigure}{0.20\linewidth}
    \centerline{\includegraphics[width=\columnwidth]{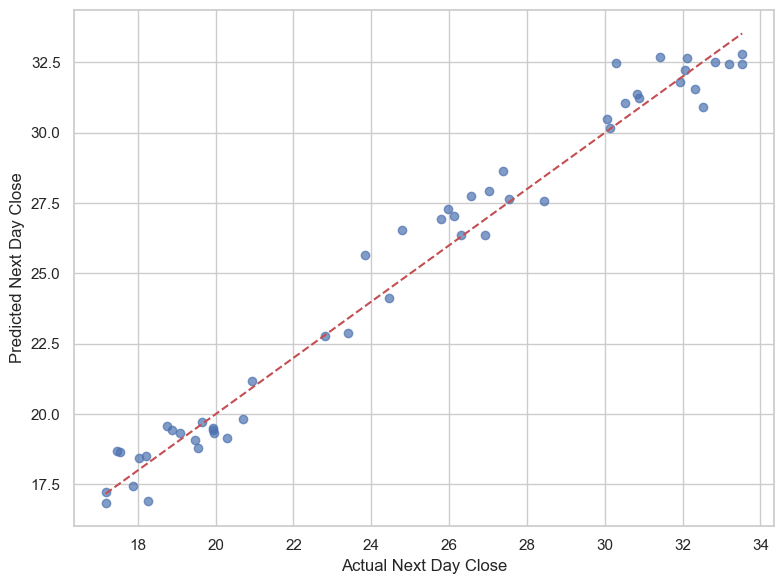}}
    \caption{Predicated V.S. Actual Prices on MTDR}
    \label{er-np-MTDR}
  \end{subfigure}
\caption{Comparison between predicated and actual next day closing prices on 8 datasets.} 
  \label{er-np}
\end{figure*}

As shown in Fig.~\ref{er-np}, visualization of predicted versus actual closing prices directly assess the model’s forecasting accuracy. For each dataset, SPH-Net’s predicted stock prices closely align with the actual values, with minimal lag or divergence, even in volatile market conditions. This confirms that the model effectively captures both short-term fluctuations and long-term patterns. The smoothness and stability of the predicted curves also demonstrate the robustness of the co-attention mechanism in handling noisy and non-linear financial data. These plots visually validate the strong regression performance shown in quantitative metrics such as R$^{2}$ and MSE.

\begin{figure*}[ht]
  \centering
  \begin{subfigure}{0.20\linewidth}
    \centerline{\includegraphics[width=\columnwidth]{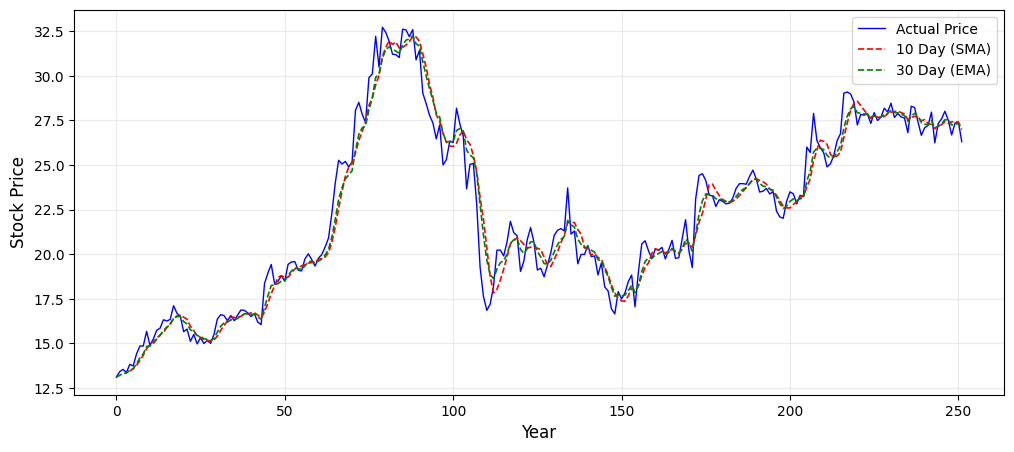}}
    \caption{Moving V.S. Actual Stock Prices on AMD}
    \label{er-v-AMD}
  \end{subfigure}
  \hfill
  \begin{subfigure}{0.20\linewidth}
    \centerline{\includegraphics[width=\columnwidth]{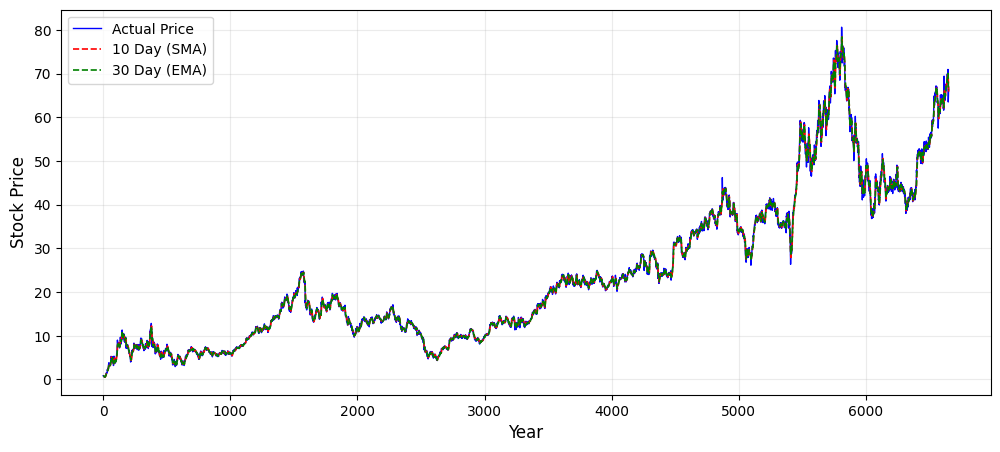}}
    \caption{Moving V.S. Actual Stock Prices on Ebay}
    \label{er-v-Ebay}
  \end{subfigure}  
  \hfill
  \begin{subfigure}{0.20\linewidth}
    \centerline{\includegraphics[width=\columnwidth]{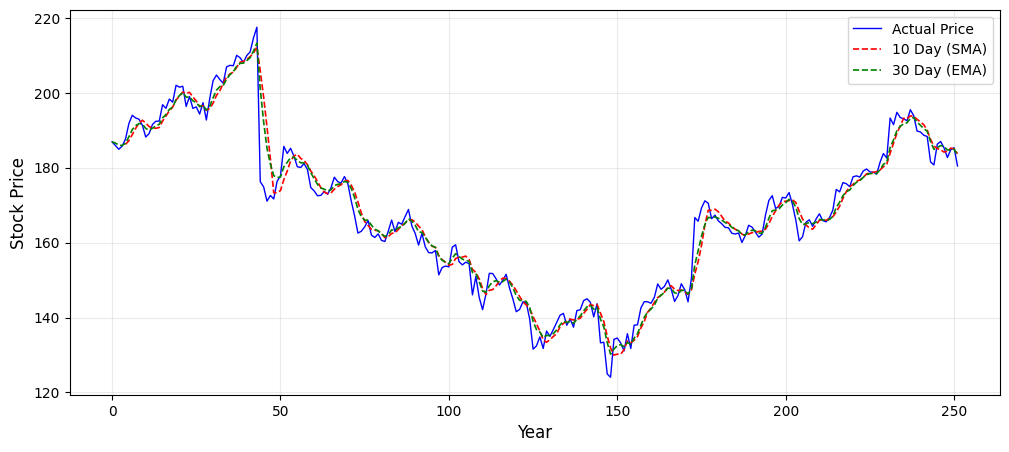}}
    \caption{Moving V.S. Actual Stock Prices on FB }
    \label{er-v-FB}
  \end{subfigure}
  \hfill
  \begin{subfigure}{0.20\linewidth}
    \centerline{\includegraphics[width=\columnwidth]{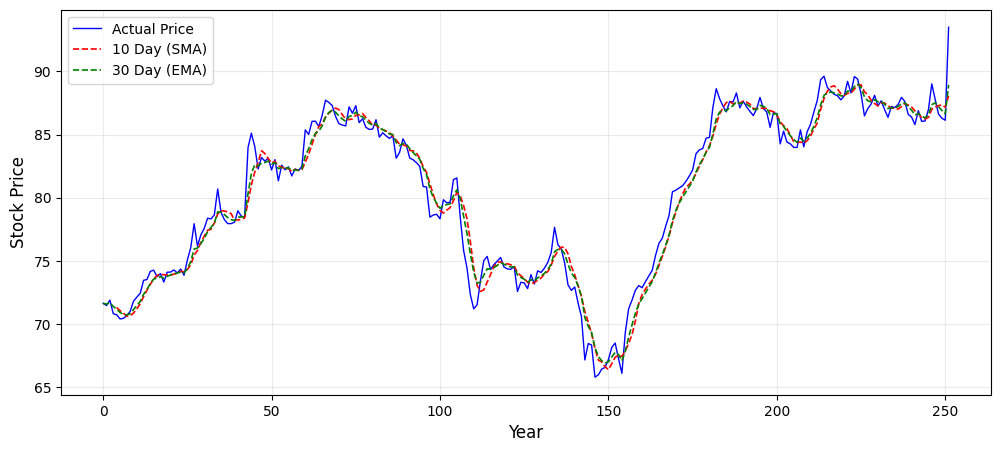}}
    \caption{Moving V.S. Actual Stock Prices on FSV }
    \label{er-v-FSV}
  \end{subfigure}
    \hfill
  \begin{subfigure}{0.20\linewidth}
    \centerline{\includegraphics[width=\columnwidth]{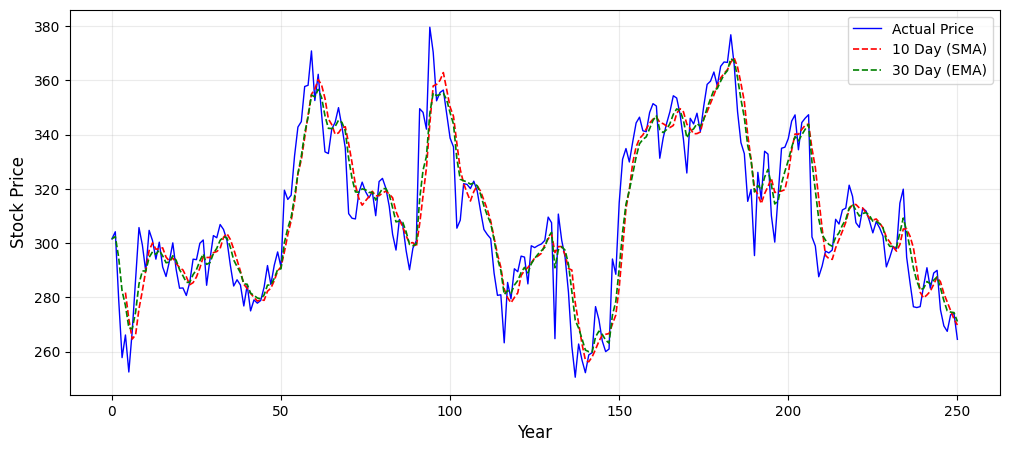}}
    \caption{Moving V.S. Actual Stock Prices on Tesla}
    \label{er-v-tesla}
  \end{subfigure}
  \hfill
  \begin{subfigure}{0.20\linewidth}
    \centerline{\includegraphics[width=\columnwidth]{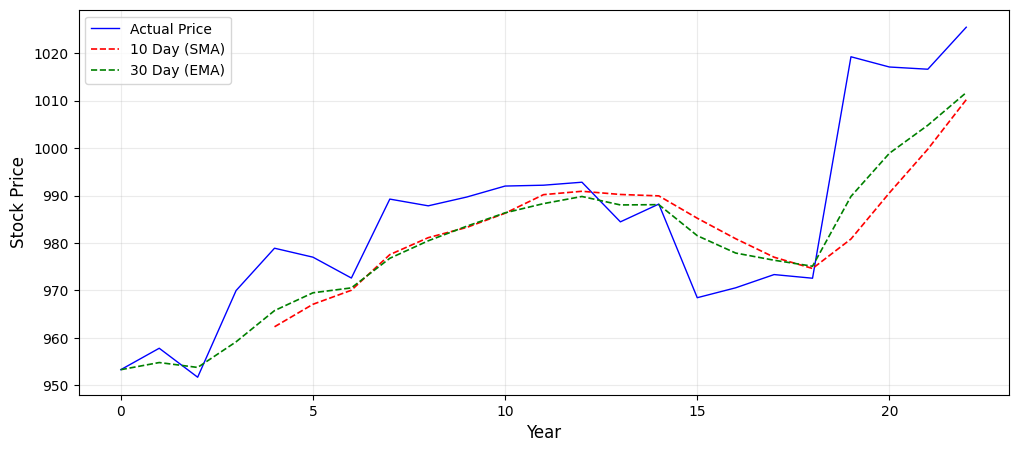}}
    \caption{Moving V.S. Actual Stock Prices on Google}
    \label{er-v-Google}
  \end{subfigure}
      \hfill
  \begin{subfigure}{0.20\linewidth}
    \centerline{\includegraphics[width=\columnwidth]{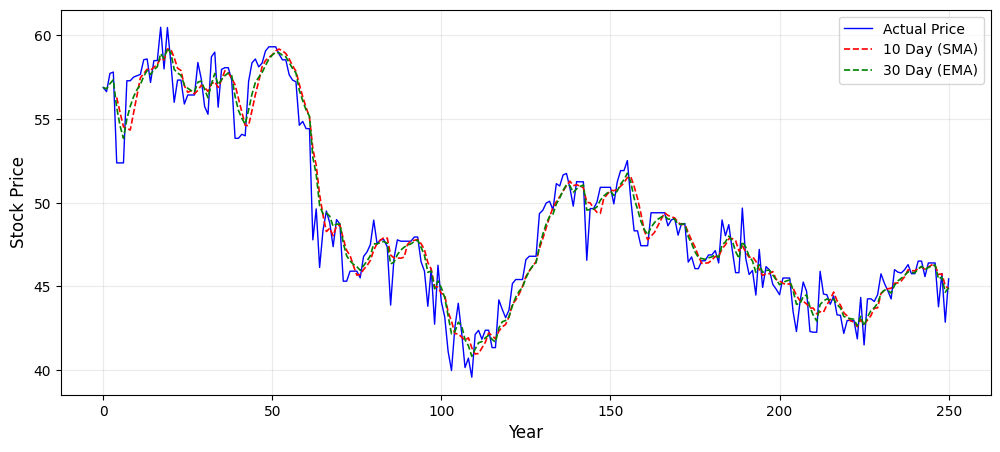}}
    \caption{Moving V.S. Actual Stock Prices on Mondy}
    \label{er-v-Mondy}
  \end{subfigure}
  \hfill
  \begin{subfigure}{0.20\linewidth}
    \centerline{\includegraphics[width=\columnwidth]{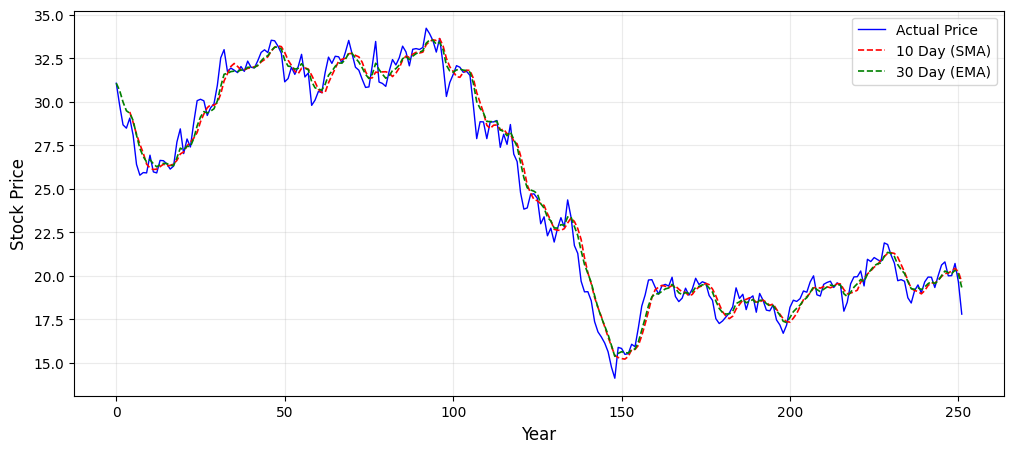}}
    \caption{Moving V.S. Actual Stock Prices on MTDR}
    \label{er-v-MTDR}
  \end{subfigure}
\caption{Moving averages V.S. actual stock prices on 8 datasets (Simple Moving Average, SMA and Exponential Moving Average, EMA).} 
  \label{er-v}
\end{figure*}

As shown in Fig.~\ref{er-v}, moving average trends (SMA and EMA) with actual stock prices can evaluate how well SPH-Net tracks underlying market trends. While moving averages provide smoothed estimates that filter out short-term noise, they often lag behind real price movements. In contrast, SPH-Net’s predictions more accurately follow the actual price trajectory while still preserving trend consistency. This indicates that SPH-Net effectively balances reactivity and stability—two crucial aspects for real-time financial forecasting. Together, these visual analyses reinforce the model's interpretability, accuracy, and applicability in real-world trading environments.

\subsection{Analysis}

The comparative experimental results demonstrate that SPH-Net achieves superior performance compared to both traditional machine learning models and recent deep learning architectures across all eight datasets. As shown in the results, the model consistently attains the lowest mean squared error (MSE) and the highest coefficients of determination R$^2$,  along with optimal precision, accuracy, and recall metrics. This performance advantage primarily originates from the synergistic integration of co-attention mechanisms and hybrid architectural design. In contrast to Transformer \cite{att} and BERT \cite{bert} architectures, SPH-Net incorporates visual feature extraction through Vision Transformer (ViT) \cite{vit}, which enables effective learning of subtle temporal patterns from raw time-series data converted into patch sequences. Notably, SPH-Net surpasses the performance of MagicNet \cite{MagicNet}, DeepClue \cite{deepclue}, and HATR-I \cite{HATR-I} models that employ textual or structural auxiliary data, thereby demonstrating that architectural innovations in SPH-Net alone provide sufficient predictive enhancement without requiring supplementary data modalities.

The ablation study systematically confirms the individual contributions of each component within SPH-Net. Experimental evidence indicates that removing the ViT module or modifying the patch size configuration results in measurable performance degradation, substantiating ViT's critical role in capturing local temporal dependencies and short-term market dynamics. The analysis further reveals that the number of attention heads in the Transformer decoder significantly impacts model effectiveness, with eight attention heads achieving an optimal balance between computational efficiency and representational capacity. Comparative experiments replacing the Transformer decoder with recurrent architectures (GRU \cite{gru}, LSTM \cite{lstm}) exhibit markedly inferior performance, thereby validating the superiority of self-attention mechanisms in modeling long-range temporal dependencies and complex inter-step interactions. These findings collectively establish that the co-attention-based integration of ViT and Transformer constitutes an essential architectural configuration for high-performance stock price prediction.

The visualization results presented in Figures~\ref{er-c}, \ref{er-np}, and \ref{er-v} provide complementary insights into SPH-Net's operational characteristics. The correlation heatmaps demonstrate the model's enhanced capability to extract discriminative features from highly correlated financial variables, a critical requirement for robust cross-domain generalization. The temporal prediction plots reveal SPH-Net's effectiveness in mitigating persistent challenges observed in traditional approaches: specifically, the elimination of prediction latency artifacts common in LSTM \cite{lstm} implementations and the avoidance of oversmoothing tendencies characteristic of ARIMA \cite{spp-arima} models. Comparative analysis with moving average trends \cite{lir,lor,gg} further demonstrates SPH-Net's improved responsiveness to rapid market fluctuations and price volatility events. These visualizations collectively confirm that the hybrid architecture - combining ViT \cite{vit} for fine-grained pattern extraction and Transformer \cite{att} for temporal dependency modeling - achieves not only numerical superiority but also produces interpretable predictions aligned with financial market dynamics, thereby fulfilling practical requirements for real-world deployment.

An interesting phenomenon worth noting is the performance of all models on Google's dataset. As shown in Fig.~\ref{gg-1}, Fig.~\ref{gg-2}, and Fig.~\ref{er-c-Google}, Google's data is particularly discrete and there is no regularity in its distribution. Fig.~\ref{er-np-Google} and Fig.~\ref{er-v-Google} are also verified, and the data generated is very scattered. Table.~\ref{er-1} and Table.~\ref{er-ab} also show a large indicator gap on MSE. Since it is just a prediction of the increase (binary classification) rather than an exact value, the models perform better in Table.~\ref{er-1-c}. Based on it, Google's dataset can be used as a good interference term to verify the robustness of the model.

\section{Conclusion}
In this paper, we introduce SPH-Net, a novel hybrid neural network with co-attention mechanisms, specifically designed for accurate stock price prediction. By integrating the Vision Transformer with a Transformer encoder-decoder architecture, SPH-Net effectively captures both local and global temporal dependencies inherent in financial time-series data. Our approach transforms conventional stock price data into a structured, image-like format, thereby facilitating more expressive feature extraction and improved trend recognition. Extensive experimentation on eight real-world stock datasets spanning multiple industries demonstrates that SPH-Net consistently outperforms state-of-the-art models across both regression and classification tasks. The model achieves superior R$^{2}$ scores, reduced MSE, and enhanced classification metrics such as accuracy, precision, and recall. Moreover, ablation studies substantiate the contributions of each architectural component, validating the efficacy of the co-attention mechanism and the hybrid structure. In conclusion, SPH-Net offers a promising framework for financial time-series forecasting, delivering enhanced prediction accuracy and robustness. Furthermore, its design is adaptable for potential extensions to multimodal data and other time-sensitive applications beyond stock market analysis.

\section{Future Work}

Despite SPH-Net's strong performance across various datasets, several avenues for future research remain. Firstly, incorporating additional data modalities, such as financial news, analyst reports, or sentiment data, could further bolster the model’s ability to capture market context and external influences. Secondly, while the current model demonstrates robust results in daily-level predictions, investigating its applicability to intra-day or high-frequency trading data presents an interesting challenge. Additionally, further advancements in the scalability and interpretability of SPH-Net are needed. Specifically, developing lightweight versions of the model for deployment in real-time systems and integrating explainable AI techniques to enhance decision-making transparency represent valuable directions for future work. Finally, extending SPH-Net to other financial forecasting tasks—such as volatility prediction or portfolio optimization—could expand its applicability in practical investment scenarios.

\end{document}